\documentclass[fleqn,usenatbib]{mnras}

\usepackage{newtxtext,newtxmath}
\usepackage[T1]{fontenc}
\usepackage{ae,aecompl}

\usepackage{subcaption}

\usepackage{xcolor}
\usepackage{graphicx}	% Including figure files
\usepackage{amsmath}	% Advanced maths commands

\usepackage{epsfig}
\usepackage{rotating}
\usepackage{color}
\usepackage{tabularx}
\usepackage{multirow}
\usepackage{graphics}
\usepackage{epsfig}
\usepackage{epstopdf}
\usepackage[para,online,flushleft]{threeparttable}
\usepackage[ampersand]{easylist}
\definecolor{navy}{rgb}{0,0.3,0.7}
\definecolor{forrest}{rgb}{0,0.55,0.25}
\definecolor{tyrianpurple}{rgb}{0.4, 0.01, 0.24}
\definecolor{coralred}{rgb}{1.0, 0.25, 0.25}
\usepackage{hyperref}
\hypersetup{
    colorlinks=true,
    linkcolor=blue,
    filecolor=blue,      
    urlcolor=blue,
    citecolor=blue,
    filecolor=blue,
}
\title[Optical and H\,{\sc i} observations of IC~443 and G189.6+3.3]{Optical and H\,{\sc i} observations of IC~443 and G189.6+3.3 in a complex environment}

\author[H. Bak{\i}\c{s} et al.]{{\color {black} H.~Bak{\i}\c{s}$^{1}$,\thanks{E-mail:{\color  {blue}hicranbakis@akdeniz.edu.tr} (HB)}
G.~Payl{\i}$^{1,2}$,
E.~Aktekin$^{3}$, 
H.~Sano$^{4}$, and A.~Sezer$^{5}$}\\
$^{1}$Department of Space Sciences and Technologies, Akdeniz University, 07058, Antalya, Turkey\\
$^{2}$Astrophysikalisches Institut und Universitätssternwarte Jena, D-07745 Jena, Germany \\
$^{3}$Department of Physics, S\"{u}leyman Demirel University, 32000, Isparta, Turkey \\
$^{4}$Faculty of Engineering, Gifu University, 1-1 Yanagido, Gifu 501-1193, Japan\\
$^{5}$Department of Computer Engineering, Avrasya University, 61250, Trabzon, Turkey\\
}

\date{Accepted XXX. Received YYY; in original form ZZZ}

\pubyear{2024}

\begin{document}
\label{firstpage}
\pagerange{\pageref{firstpage}--\pageref{lastpage}}
\maketitle

% Abstract of the paper
\begin{abstract}
The supernova remnant (SNR) IC~443 is one of the best-studied Galactic SNRs at many wavelengths. It is interacting with a very complex environment, including the SNR G189.6+3.3 and H\,{\sc ii} regions. In this paper, we observed IC~443 and G189.6+3.3 using 1.5- and 1-m telescopes to better understand the nature of these SNRs in the optical band. We perform H$\alpha$ images showing both filamentary and diffuse structures, and long-slit spectra from many locations with Balmer and forbidden lines detected for IC~443 and G189.6+3.3. The [S\,{\sc ii}]/ H$\alpha$ ratios confirm the SNR nature of G189.6+3.3. The ranges of our estimated electron density and pre-shock cloud density clearly indicate the complex structure surrounding IC~443 and G189.6+3.3. We also investigated the archival H\,{\sc i} data and newly found some shell-like distributions of H\,{\sc i} that are possibly associated with G189.6$+$3.3.

\end{abstract}


%%%%%%%%%%%%%%%%%%%%%%%%%%%%%%%%%%%%%%%%%%%%%%%%%%

%%%%%%%%%%%%%%%%% BODY OF PAPER %%%%%%%%%%%%%%%%%%

\section{Introduction}
Detecting optical emission from supernova remnants (SNRs) is critical to reveal the SNR properties and the environmental conditions (e.g. ambient density) that should have affected the SNR evolution. Optical CCD imaging with H$\alpha$, [S\, {\sc ii}] and [O\, {\sc iii}] filters, and spectroscopic observations of SNRs helped identify the SNR's nature  (e.g. \citealt{St12, Sa13, Fe15, Ho18, Bo22, Pa24, Fe24}).

IC~443 (G189.1+3.0, Sh2-248, Jellyfish Nebula) is a mixed-morphology (MM; \citealt{RhPe98}) SNR interacting with surrounding molecular clouds (MCs) and has been well studied from radio to gamma-rays (e.g. \citealt{Yam09, Ab10, Zh18}). It is an evolved SNR (age of $3-30$ kyr: \citealt{Pe88, Ol01}) located in the Galactic anti-centre at a distance range 1.5 to 2 kpc \citep{Pe88, WeSa03, AC17}. The extinction distance to the SNR was estimated to be 1.7$-$1.8 kpc (see \citealt{Yu19, Zh20}). 

Previous studies of IC~443 mainly focused on the bright northeastern (NE) rim and the filaments extending beyond ($\sim$15 arcmin) the NE rim to the northeast (e.g. \citealt{Fe84, Br86}). The radio bright NE rim is evolving in a nonuniformly medium with a non-circular morphology and two shells, namely Shell A and B (e.g. \citealt{Br86}). \citet{Le08} analyzed 21 cm spectral-line and continuum observations using the Very Large Array (VLA) and the Arecibo telescope. They showed that Shell A has a nearly circular rim, and is located in the NE region. Shell B is located in the southwestern (SW) region and shows a larger and more extended structure with an apparent breakout associated with evolution into a rarefied medium. \citet{Br86} defined a third shell (Shell C), which is faint and extends beyond the NE region. Analyzing the {\it ROSAT} All-Sky Survey data,  \citet{As94} proposed that Shell C is a new SNR called G189.6+3.3. 

Recent observational (e.g. \citealt{De20, Co22}) and theoretical (e.g. \citealt{Us21}) studies have examined the shock interaction between IC~443 and the nearby molecular clump G.

In the optical band, the NE rim was investigated in several studies (e.g. \citealt{Do74, Fe80, AC17, AlDr19}). \citet{Fe80} obtained spectra from six locations in the NE rim and estimated the electron densities ranging from less than 100 up to 500 cm$^{-3}$. They found the shock velocities of 65$-$100 km s$^{-1}$, with a pre-shock number density of  
10$-$20 cm$^{-3}$. \citet{AC17} presented the kinematics of the NE rim using a scanning Fabry–Perot interferometer at the H$\alpha$ line. They estimated a distance of 1.9 kpc to the SNR and of 1.3 kpc to the H\,{\sc ii} region S249. \citet{AlDr19} performed optical observations of the NE rim using the Canada-France-Hawaii Telescope imaging spectrograph SITELLE. They found that extinction shows significant variation across the observed region with $E(B-V)$ = 0.8$-$1.1, and electron densities range between 100 and 2500 cm$^{-3}$.

Using the 1.3 m telescope at McGraw-Hill Observatory, \citet{Fe84} observed the relatively faint NE filaments, which are located about 15 arcmin east of the NE rim and along the southern boundary of the H\,{\sc ii} region S249.  The author performed optical spectrophotometry of three locations in the NE filaments and two locations in the H\,{\sc ii} region S249.  They examined an association between NE filaments and IC~443, and showed that the NE filaments represent shock-heated gas.

SNR G189.6+3.3 was first discovered in X-rays with the {\it ROSAT} All-Sky Survey \citep{As94}. The X-ray image shows a ring-like morphology with a diameter of $\sim$1.5 arcmin. The centre of this SNR is offset from the centre of IC~443 by $\sim$0$\fdg$7.  X-ray spectrum of the SNR showed that the emission is soft with an electron temperature of $\sim$0.14 keV. This result suggested that G189.6+3.3 is an evolved SNR (age of $\sim$ $10^{5}$ yr). They estimated a distance to SNR of $d$ $\sim$ 1.5 kpc. \citet{Ya20} analyzed the {\it Suzaku} data of the NE region of G189.6+3.3, and concluded that it is most likely to be a middle-aged SNR with a recombining plasma. Recently, \citet{Ca23} presented a full X-ray spectral characterization of G189.6+3.3 emission using the data provided by {\it eROSITA} on board the Spectrum Roentgen Gamma mission. They confirmed that G189.6+3.3 is a SNR and concluded that G189.6+3.3 completely overlaps with IC~443. The authors proposed that the progenitors of G189.6+3.3 and IC~443 could have been expelled from the same binary or multiple systems.

Although IC~443 is one of the best-studied SNRs, the nature of the emission from G189.6+3.3 has not been fully understood. Previous studies provoked some interesting questions about these SNRs: (i) Whether G189.6+3.3 and IC~443 are related or not, (ii) Are they two separate SNRs?, and (iii) Could they occur from the same supernova mechanism? The local ambient properties are important to investigate these questions. Therefore, aiming at exploring the properties of these SNRs and their ambient environment, we perform optical photometric and spectroscopic observations and analyze the archival H\,{\sc i} data of IC~443 and G189.6+3.3 regions.

In this work, we report the detection of optical emission from G189.6+3.3 showing both filamentary and diffuse structure, present new H$\alpha$ images and spectra of IC 443, and found some shell-like distributions of H\,{\sc i} that are possibly associated with G189.6$+$3.3. Below, we first describe our observations and data reduction in Section \ref{obs}. We present the optical and H\,{\sc i} analyses and results in Section \ref{analysis}. We then discuss the implications of our findings and summarize the conclusions in Section \ref{discuss}.

\section{Observations and data reduction}
\label{obs}
\subsection{Optical}
We performed the photometric observations of IC~443 and G189.6+3.3  with the 1.0 m fully automatic Ritchey-Chrétien T100 telescope of T\"{U}B\.{I}TAK National Observatory (TUG)\footnote{\url{https://tug.tubitak.gov.tr/en}}, in Turkey between 2020 and 2022. For imaging, we used the CCD camera consists of $4096\times4096$ pixels, each of 15 ${\mu}$m $\times$ 15 ${\mu}$m, covering $21.5 \times 21.5$ arcmin$^2$ field of view (FoV). We observed several regions of IC~443 and G189.6+3.3 with T100 telescope using H$\alpha$ and continuum filters. In addition, only one photometric image was taken with the RTT150 telescope (see \citealt{Ba23} for focal plane instruments), which is also located at TUG and where we made the spectral observations. The log of our photometric observations is illustrated in Table \ref{Table1}. Exposure times varied from 120s to 8100s depending on sky conditions.

\begin{table*}
 \caption{Summary of photometric observations.}
 \begin{tabular}{@{}p{1.5cm}p{1.5cm}p{1.5cm}p{1.4cm}p{1.9cm}p{1.4cm}@{}}
 \hline
Region ID &  R.A. & Dec.  &   Exposure & Observation  & Appears in \\
 & (J2000) & (J2000) &  (s) & date  &  figure   \\

\hline
1 & 06:18:18 & +23:17:41  & 5 $\times$ 200 & 2021 Nov 04 &  Fig. \ref{figure1}  \\

2 & 06:22:36 & +23:14:28  & 19 $\times$ 150 & 2021 Nov 04 &  Fig. \ref{figure1} \\

3 & 06:17:27 & +22:49:45  & 21 $\times$ 150 & 2021 Nov 03 &  Fig. \ref{figure3}  \\

4 & 06:15:53 & +22:42:49  & 9 $\times$ 900 & 2020 Nov 17 &  Fig. \ref{figure3}  \\

5 & 06:16:38 & +21:57:10  & 4 $\times$ 500 & 2020 Oct 15 &  Fig. \ref{figure5}  \\

6 & 06:16:18 & +22:31:15  & 1 $\times$ 900 & 2020 Oct 15  &  Fig. \ref{figure5}  \\

7 & 06:20:29 & +22:49:14  & 1 $\times$ 600 & 2020 Nov 12 &   Fig. \ref{figure7}   \\

8 & 06:18:57 & +22:38:47  &  1 $\times$ 600 & 2020 Nov 12 &   Fig. \ref{figure7}   \\

9 & 06:18:37  & +21:54:53   & 2 $\times$ 600 & 2022 Dec 23 &  Fig. \ref{figure8}    \\
10 & 06:24:35 &  +22:57:59   & 2 $\times$ 600  & 2022 Oct 25 &  Fig. \ref{figure8} \\ 
11 & 06:22:00 & +23:00:00  & 1 $\times$ 120  & 2023 Mar 27 &  Fig. \ref{figure8a}\\ 
 \hline
 \hline
Telescope &Filter       			&    & Wavelength     &	 & FWHM             \\ 
   &          			&   & (nm)	             &  &  (nm)             \\ 
 \hline
T100&H$\alpha$     		    &    & 656.0           &   & 10.8                \\
&Cont blue                   &    &  551.0          &  &  88                     \\  
RTT150 & H$\alpha$    		    &    & 656.3          &   & 5.0              \\

\hline
\label{Table1}
\end{tabular}
\end{table*}

We obtained long-slit spectra using the Ritchey-Chrétien RTT150 1.5-m telescope with the TFOSC (TUG Faint Object Spectrograph and Camera)\footnote{\url{https://tug.tubitak.gov.tr/en/teleskoplar/rtt150-telescope-0}} instrument. We used grism number 15 has a spectral coverage of 3230$-$9120 {\AA} with a 134 $\mu$m-wide slit. For our analysis, we selected the wavelength range 4800$-$6800 {\AA}, which includes the H$\beta$, [O\,{\sc iii}], [O\,{\sc i}], H$\alpha$, [N\,{\sc ii}], and [S\,{\sc ii}] emission lines. A summary of the spectroscopic observations is given in Table \ref{Table2}. Exposure times for the spectroscopic observations varied from 1800 to 2$\times$1800s depending on sky conditions. We also observed Iron–Argon (Fe-Ar) lamp to perform the wavelength calibration. To the flux calibration, the spectrophotometric standard stars Feige 34, BD+28D4211, and  BD+75d325 from \citet{Ok90} were observed during the same nights as the SNRs. 

Standard data reductions of the images and spectra were performed using {\sc Image Reduction Analysis Facility} ({\sc iraf})\footnote{\url{https://iraf-community.github.io/}} software. This included bias and dark frame subtraction, flat-field division, and bad-pixel corrections for the images. The reduction of the spectra consisted of bias, flat-field, background-light corrections, and flux and wavelength calibrations. We also extracted the atmospheric molecular lines and city lights effects from the spectra.

\begin{table*}
\centering
 \caption{Summary of spectroscopic observations.}
 \begin{tabular}{@{}p{1.6cm}p{1.6cm}p{2.2cm}p{1.8cm}p{2.0cm}@{}}
 \hline
Slit ID & R.A.  & Dec.  & Exposure & Observation \\
 &  (J2000) & (J2000) &  (s) & date \\
\hline
3m  &06:17:40.97  & +22:49:33.45   & 2 $\times$ 1800 & 2022 Dec 22   \\
3p  &06:17:40.56  & +22:51:55.48   & 2 $\times$ 1800 & 2022 Dec 22  \\
3s  &06:17:27.26 & +22:51:55.54   & 2 $\times$ 1800 & 2022 Dec 22  \\
3b  & 06:18:02.67  & +22:46:55.38   & 2 $\times$ 1800 & 2023 Jan 23  \\
3c  & 06:18:01.45 & +22:46:55.38   & 2 $\times$ 1800 & 2023 Jan 23 \\
3e  & 06:18:07.88 & +22:47:30.95   & 2 $\times$ 1800  & 2023 Jan 23 \\
3f   &06:17:56.27  & +22:47:32.83   & 2 $\times$ 1800 & 2023 Jan 23  \\
3g  & 06:17:26.99  & +22:52:30.62   & 2 $\times$ 1800 & 2023 Jan 23 \\
3h  & 06:17:30.93 & +22:52:30.64   & 2 $\times$ 1800  & 2023 Jan 23 \\
3i  & 06:17:36.98  & +22:52:30.65   & 2 $\times$ 1800 & 2023 Jan 23 \\
3j   & 06:17:38.19  & +22:52:30.65  & 2 $\times$ 1800 & 2023 Jan 23 \\
3l   & 06:18:17.28  & +22:41:45.53  & 2 $\times$ 1800 & 2023 Jan 23  \\ [0.5 ex]
\hline
7a  &06:19:53.05  & +22:45:48.72   & 2 $\times$ 1800   &   2022 Dec 21   \\
7b  &06:19:48.99  & +22:45:47.77   & 2 $\times$ 1800   &  2022 Dec 21    \\
7c  & 06:19:44.05  & +22:46:08.18   & 1 $\times$ 1800 & 2023 Mar 27  \\
7d  & 06:19:53.68  & +22:46:08.21   & 1 $\times$ 1800 & 2023 Mar 27  \\
7e  & 06:19:51.38  & +22:46:06.80 & 2 $\times$ 1800 & 2023 Mar 27 \\
7f  & 06:20:05.27  & +22:45:56.06  &  2 $\times$ 1800 & 2023 Mar 27   \\
7g  & 06:19:57.88  & +22:45:50.35  &  2 $\times$ 1800 & 2023 Mar 27  \\ [0.5 ex]
\hline
4a  &06:16:09.86  & +22:42:30.64   & 2 $\times$ 1800  &  2022 Dec 21   \\
4b  &06:16:05.12 & +22:42:29.60   & 2 $\times$ 1800   & 2022 Dec 21    \\  
11a  & 06:21:49.21 & +22:59:42.08   & 1 $\times$ 1800 & 2023 Mar 27 \\ [0.5 ex]
\hline
 \hline
\label{Table2}
\end{tabular}
\end{table*}

\subsection{H\,{\sc i}}
We used archival data of the H\,{\sc i} 21~cm line to reveal the neutral gas distributions surrounding the SNRs IC~443 and G189.6$+$3.3. The H\,{\sc i} data are from the Canadian Galactic Plane Survey (CGPS; \citealt{Ta03}), which were taken at the Dominion Radio Astrophysical Observatory (DRAO). The angular resolution is 58~arcsec $\times$ 152~arcsec toward the region. The typical noise fluctuation of H\,{\sc i} is $\sim$3~K at the velocity resolution of 0.82~km~s$^{-1}$.

\section{Analysis and Results}
\label{analysis}
\subsection{Optical analysis and results}
\subsubsection{Images}
\begin{figure*}
\includegraphics[angle=0, width=11.5cm]{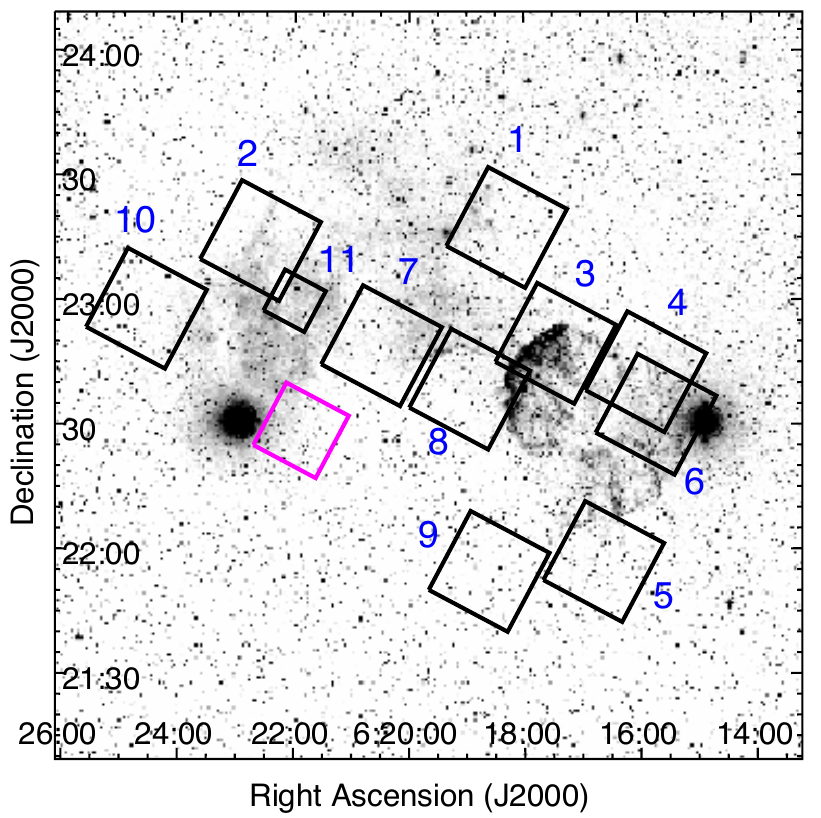}
\caption{Optical image of IC~443 and G189.6+3.3 regions obtained from the DSS  covering a field of view (FoV) of 3 $\times$ 3 deg$^{2}$. The black boxes show the regions we observed with a 1.0-m telescope (Regions 1$-$10) and RTT150 telescope (Region 11). The boxes coordinates are given in Table \ref{Table1}. The magenta box indicates the FoV of {\it Suzaku} observation.}
\label{radio}
\end{figure*}

The Sky View Virtual Observatory (DSS-Digitized Sky Survey) image of IC~443 and G189.6+3.3 regions taken from the DSS database is given in Fig. \ref{radio}. The positions of eleven regions we detected optical emission are shown in this image. 

\begin{figure*}
\includegraphics[angle=0, width=9.3cm]{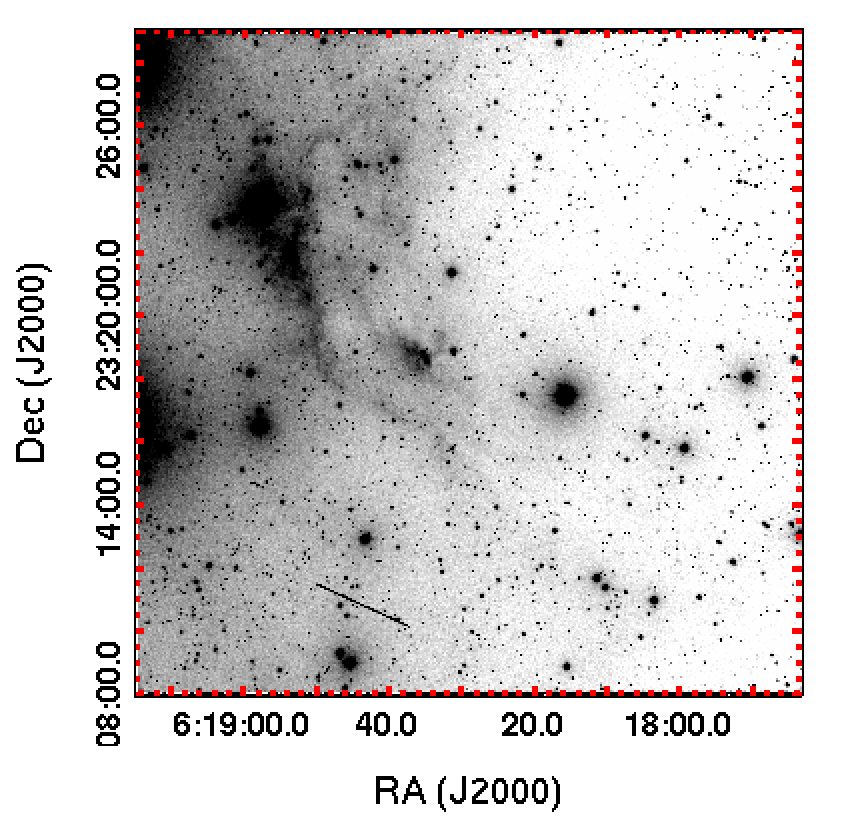}
\includegraphics[angle=0, width=8cm]{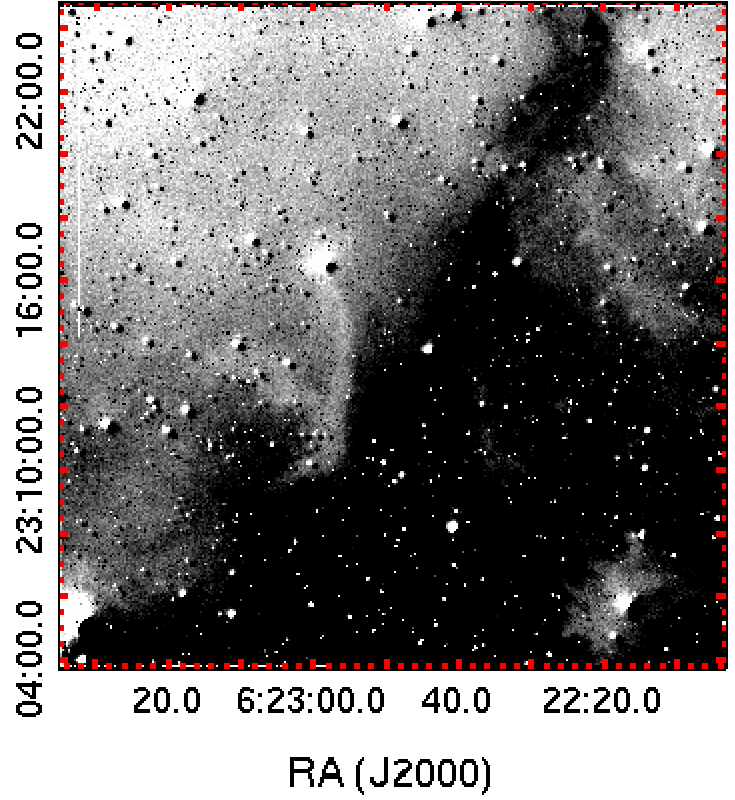}
\caption{The H$\alpha$ image of the Region 1 and the continuum-subtracted H$\alpha$ image of Region 2. North is up, East is to the left.}
\label{figure1}
\end{figure*}

\begin{figure*}
\includegraphics[angle=0, width=8.3cm]{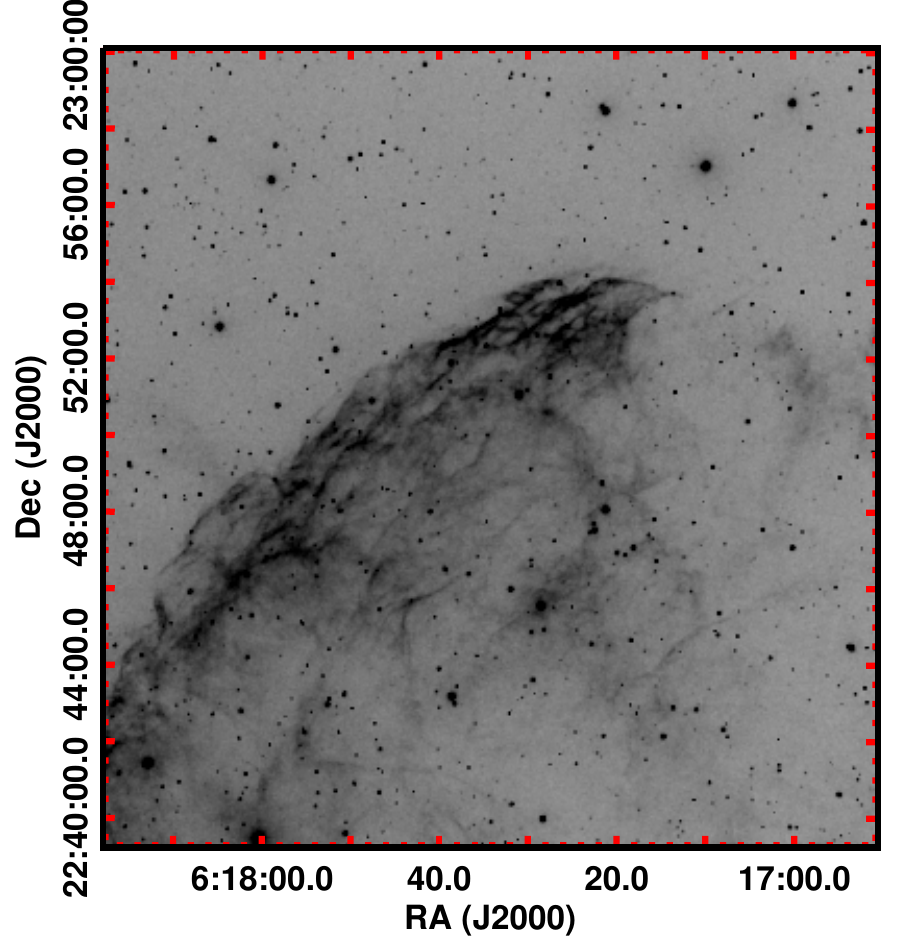}
\includegraphics[angle=0, width=8.4cm]{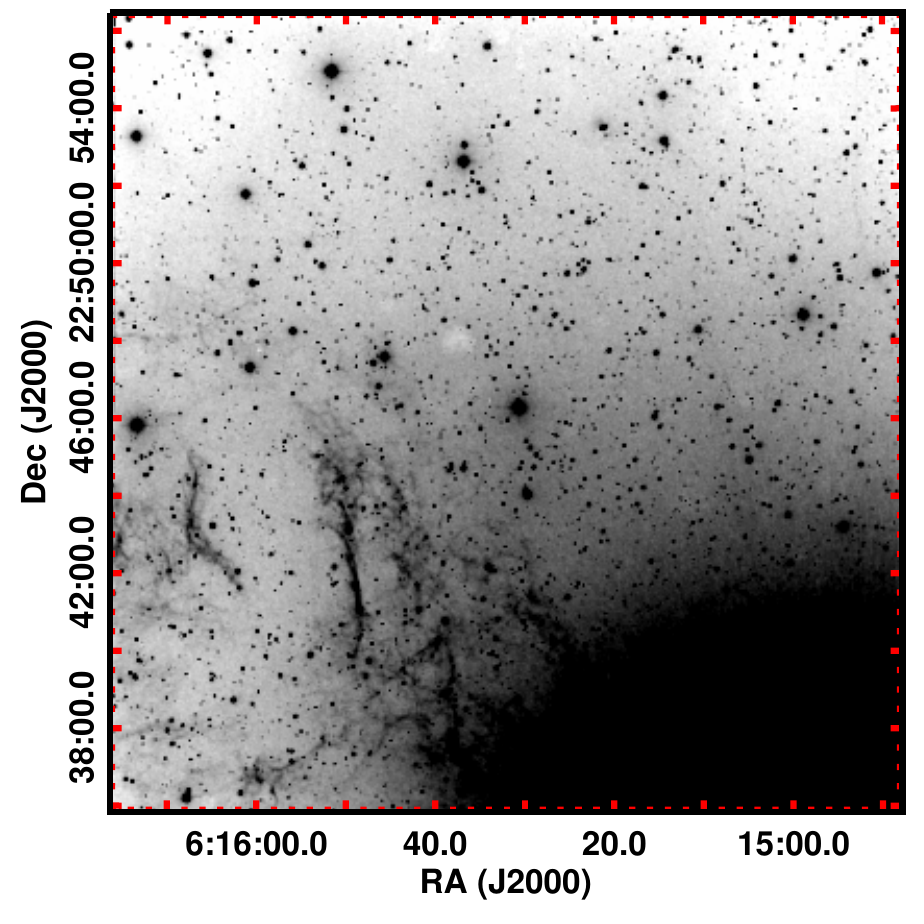}
\caption{The H$\alpha$ image of the Regions 3 and 4.}
\label{figure3}
\end{figure*}

\begin{figure*}
\includegraphics[angle=0, width=9.1cm]{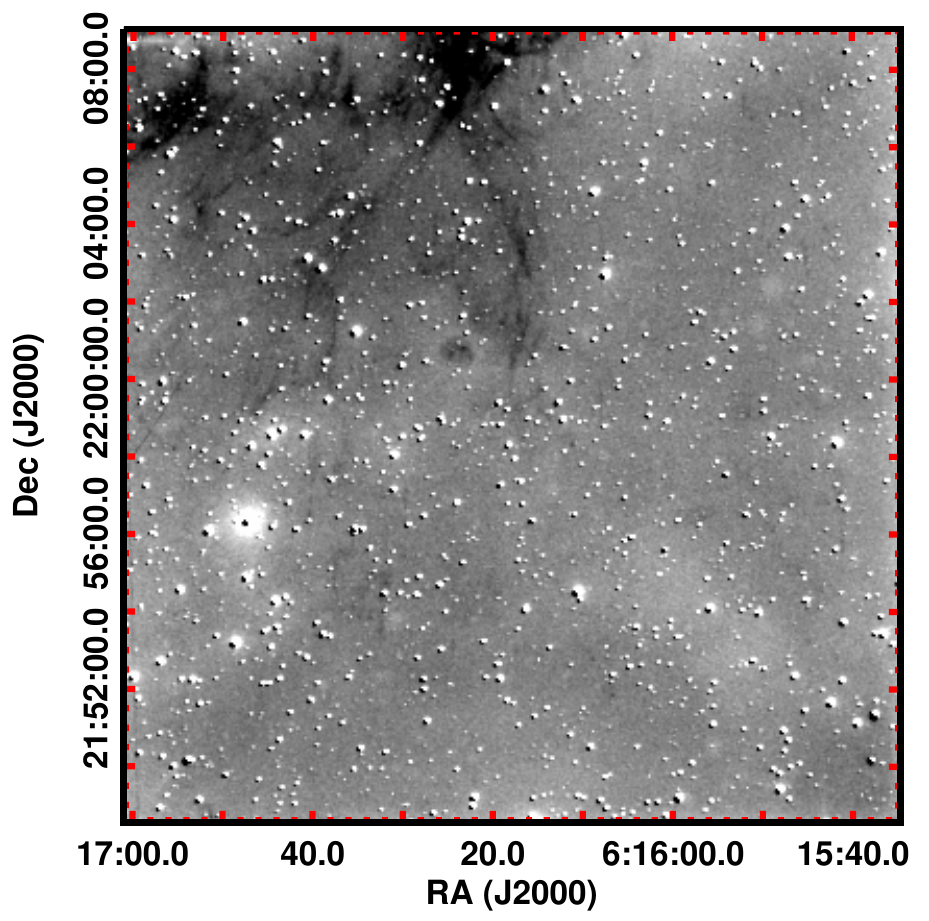}
\includegraphics[angle=0, width=8.5cm]{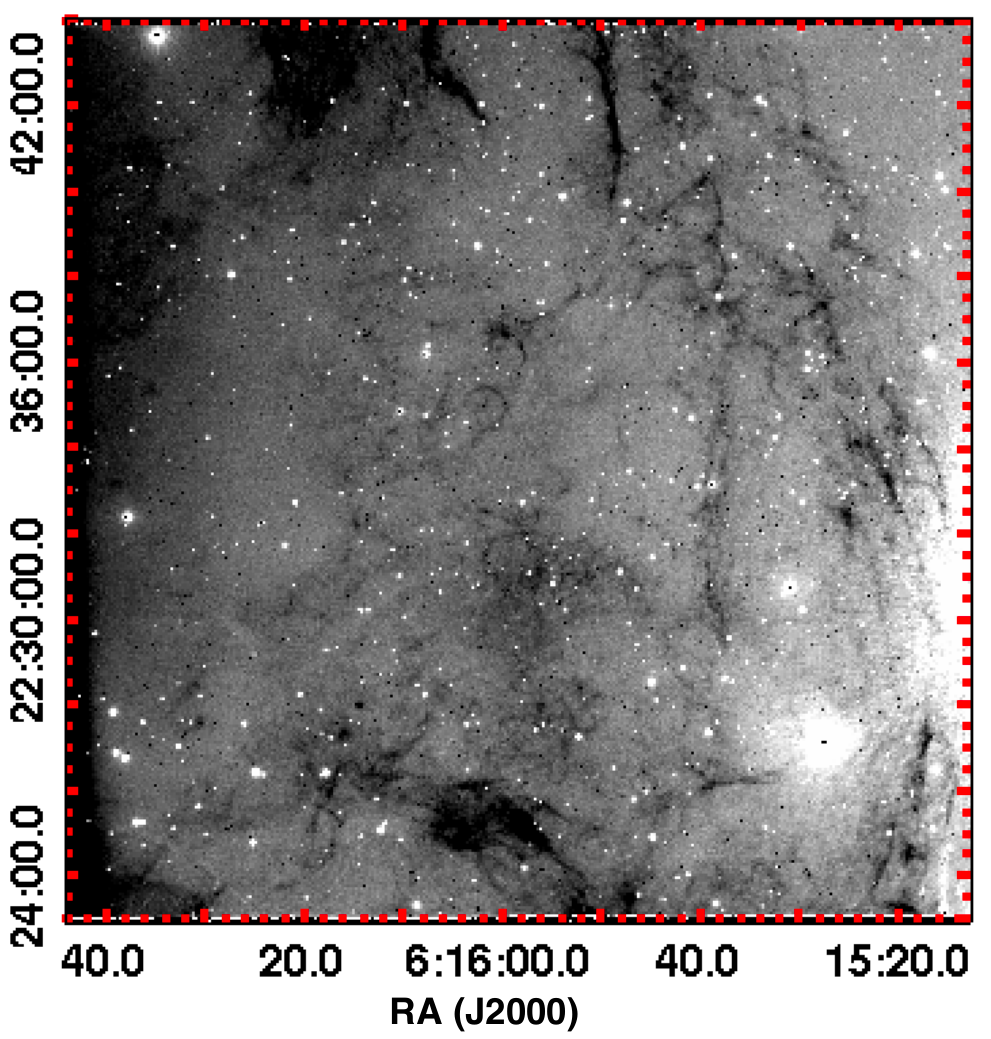}
\caption{The continuum-subtracted H$\alpha$ image of the Regions 5 and 6.}
\label{figure5}
\end{figure*}

\begin{figure*}
\includegraphics[angle=0, width=9.0cm]{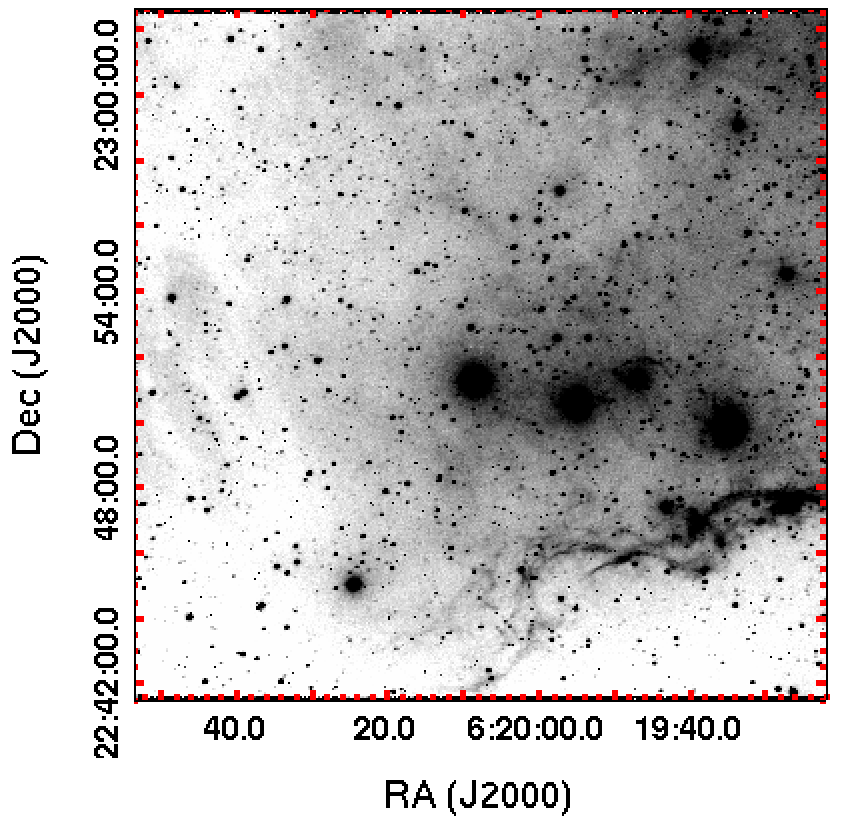}
\includegraphics[angle=0, width=8.1cm]{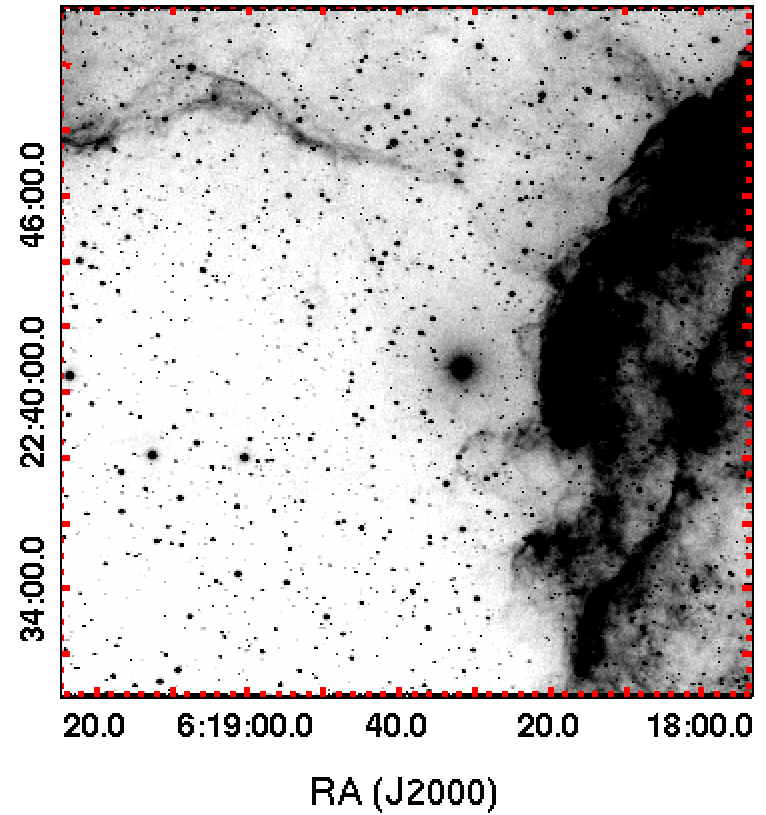}
\caption{The H$\alpha$ image of the Regions 7 and 8.}
\label{figure7}
\end{figure*}

\begin{figure*}
\includegraphics[angle=0, width=8.9cm]{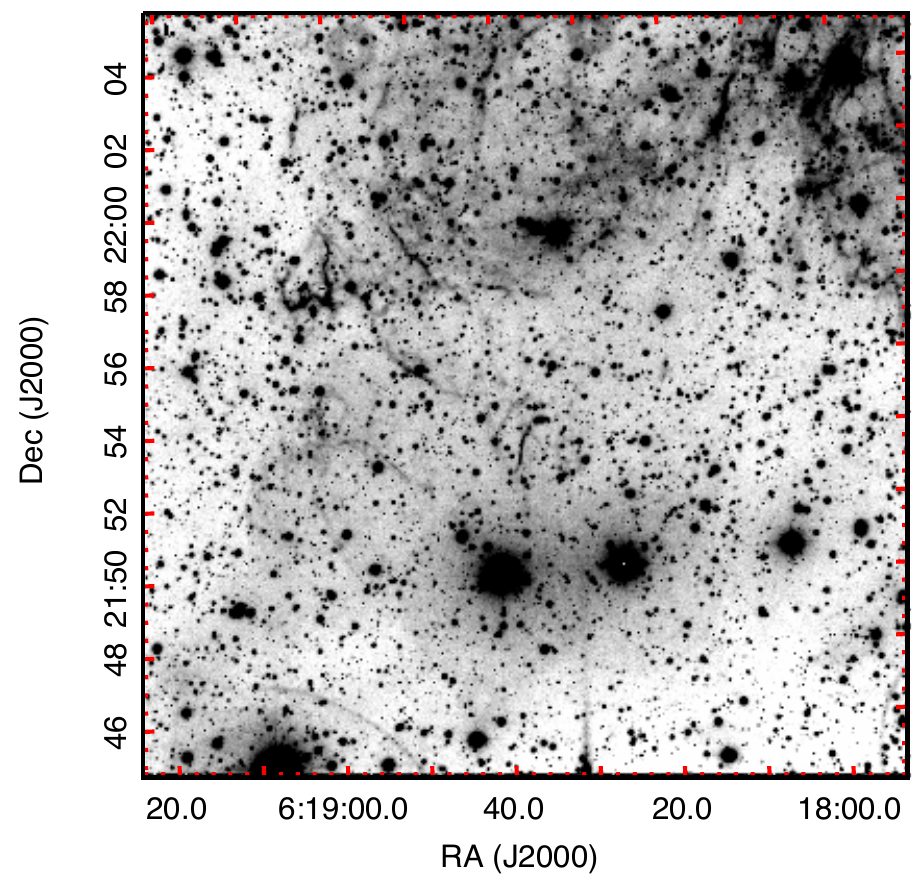}
\includegraphics[angle=0, width=8.0cm]{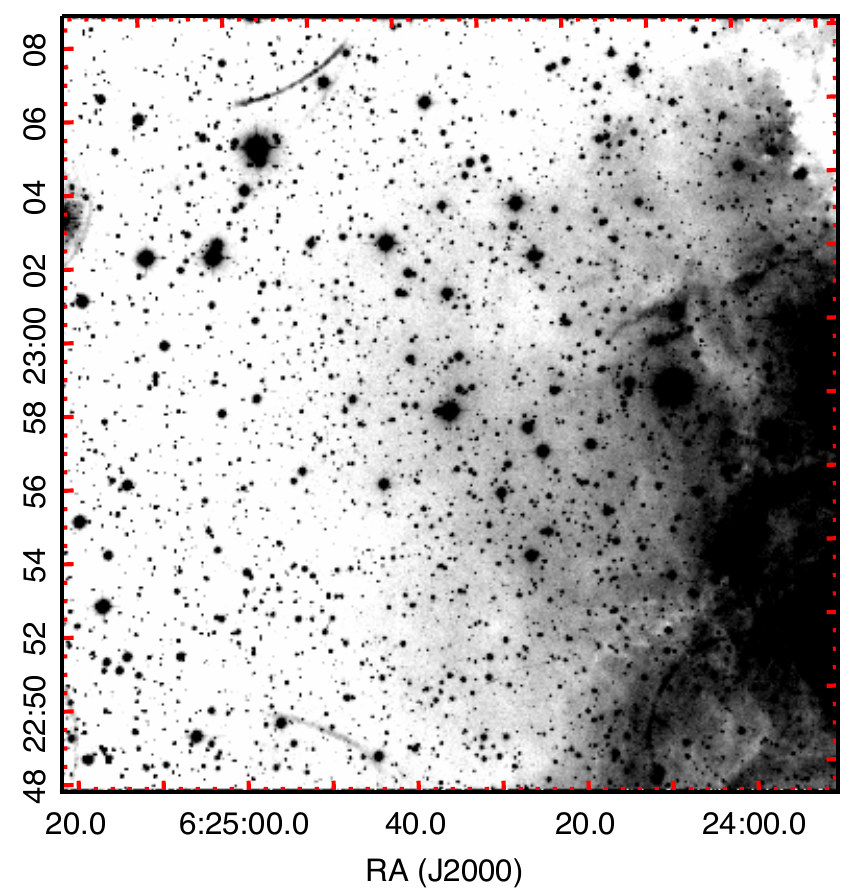}
\caption{The H$\alpha$ image of the Regions 9 and 10.}
\label{figure8}
\end{figure*}

We also presented H$\alpha$ images of these regions in Figs \ref{figure1}$-$\ref{figure8a},  where the diffuse and filamentary structure of the emission is shown clearly.  In comparison to the DSS R-map of these regions, the long curved, several thin and partial filaments, and several bright diffuse emissions are clearly resolved in our new H$\alpha$ images shown in Figs \ref{figure1}$-$\ref{figure8a}. 

\subsubsection{Spectra}
\begin{figure*}
\includegraphics[angle=0, width=7.2cm]{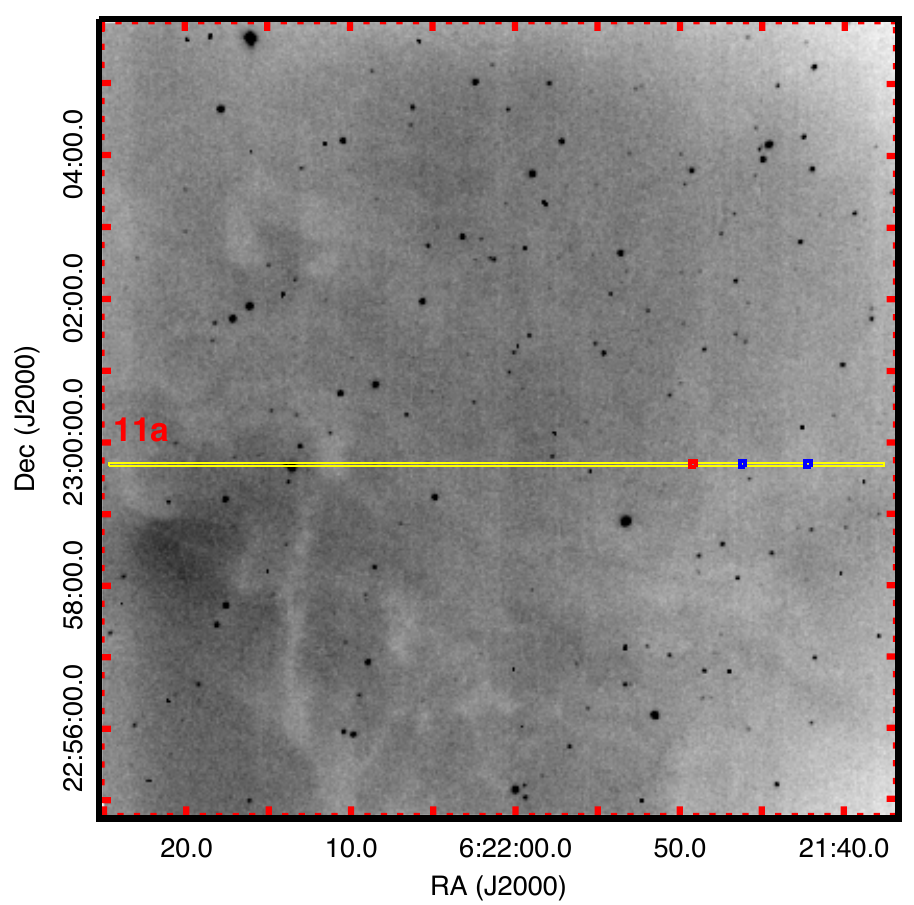}
\includegraphics[angle=0, width=7.5cm]{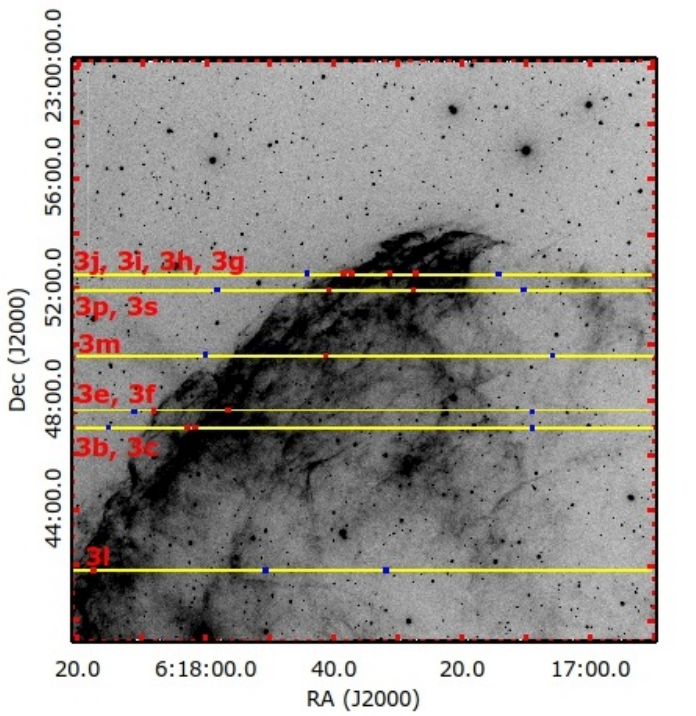}
\includegraphics[angle=0, width=7.0cm]{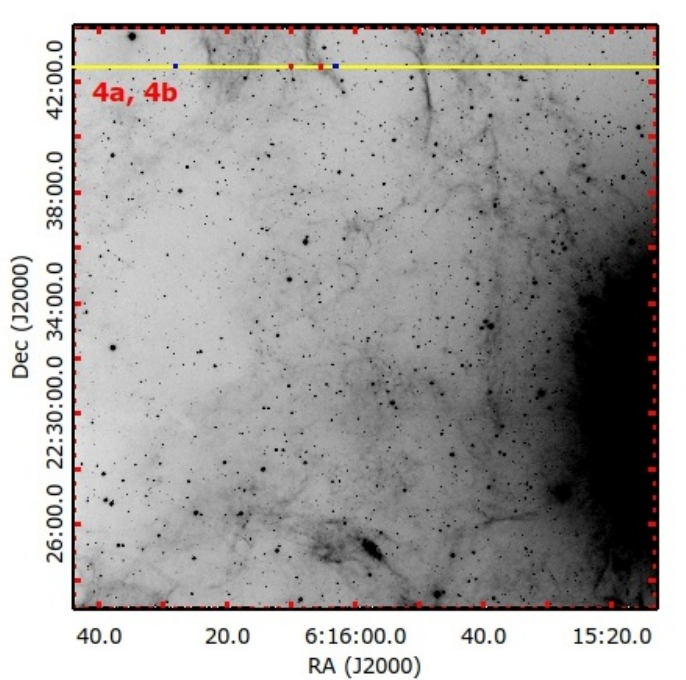}
\includegraphics[angle=0, width=7.6cm]{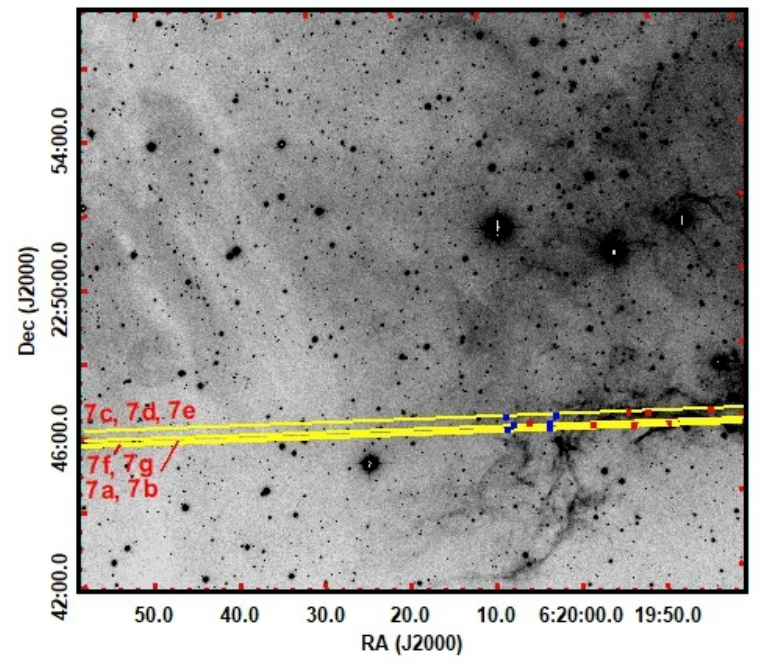}
\caption{The locations of the slits are shown by the yellow lines in the figures. The red boxes correspond to the regions where the SNR spectra were extracted, and the blue boxes correspond to the locations for the sky subtraction along each slit.}
\label{figure8a}
\end{figure*}

\begin{figure*}
\includegraphics[angle=0, width=8.5cm]{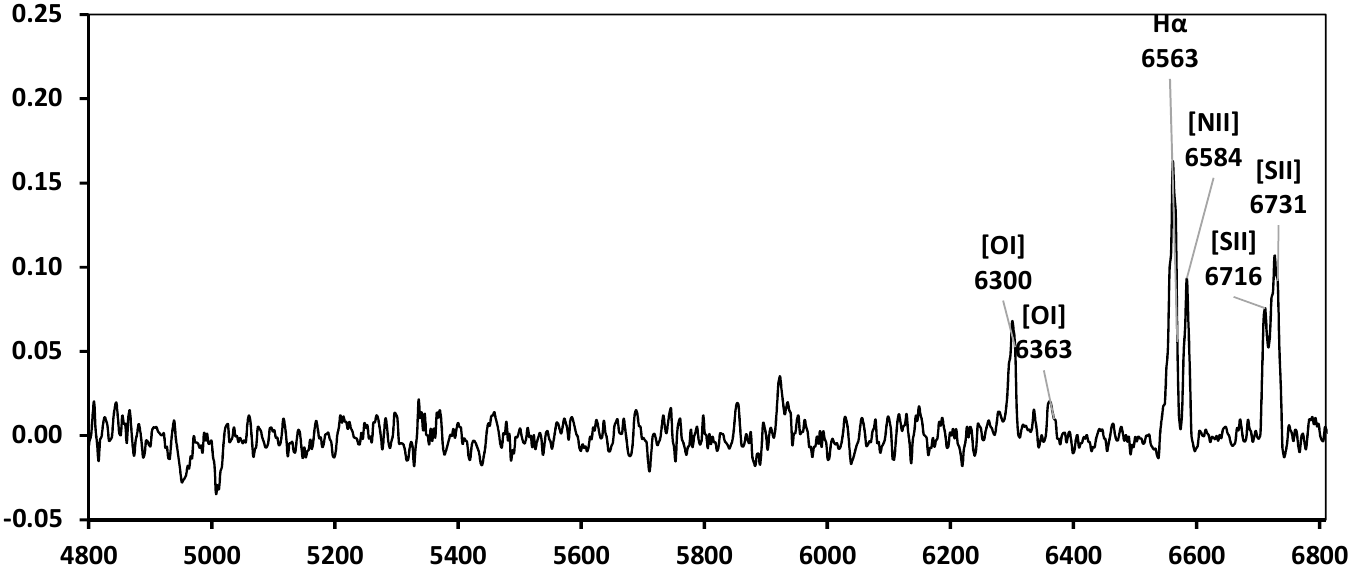}
\includegraphics[angle=0, width=8.5cm]{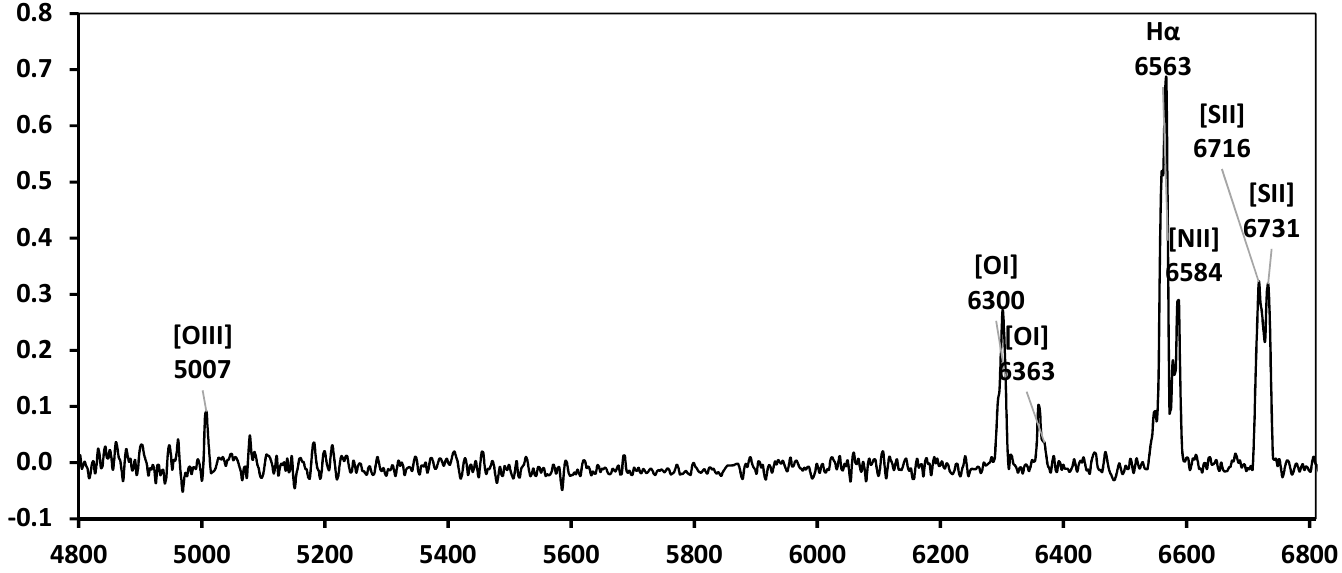}
\includegraphics[angle=0, width=8.5cm]{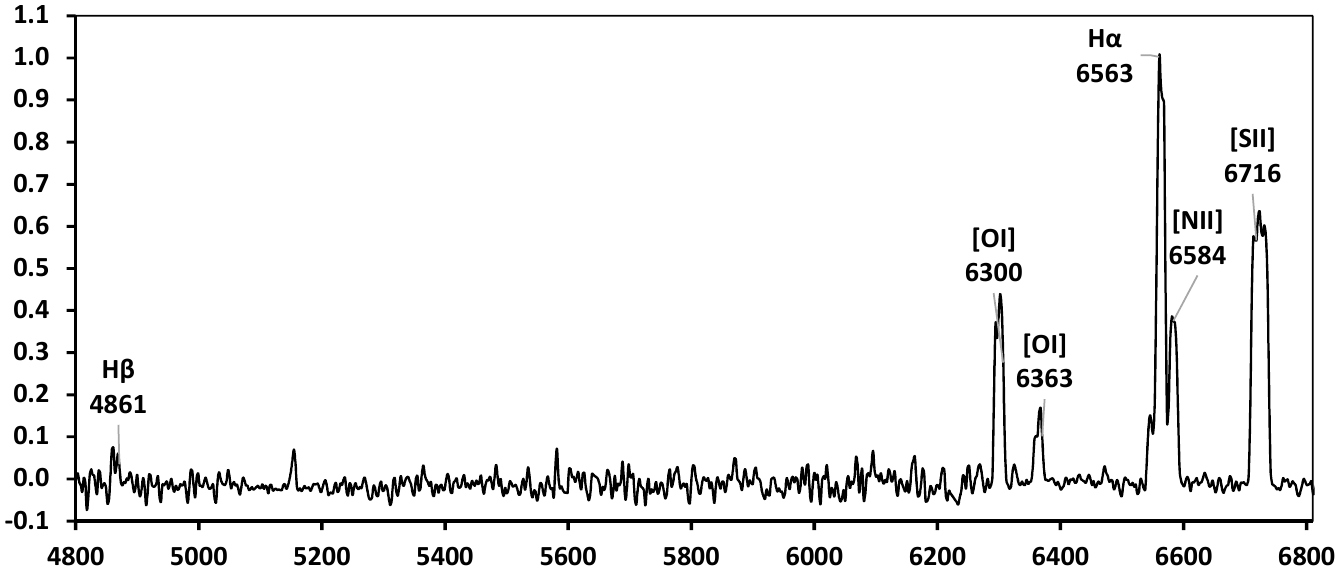}
\includegraphics[angle=0, width=8.5cm]{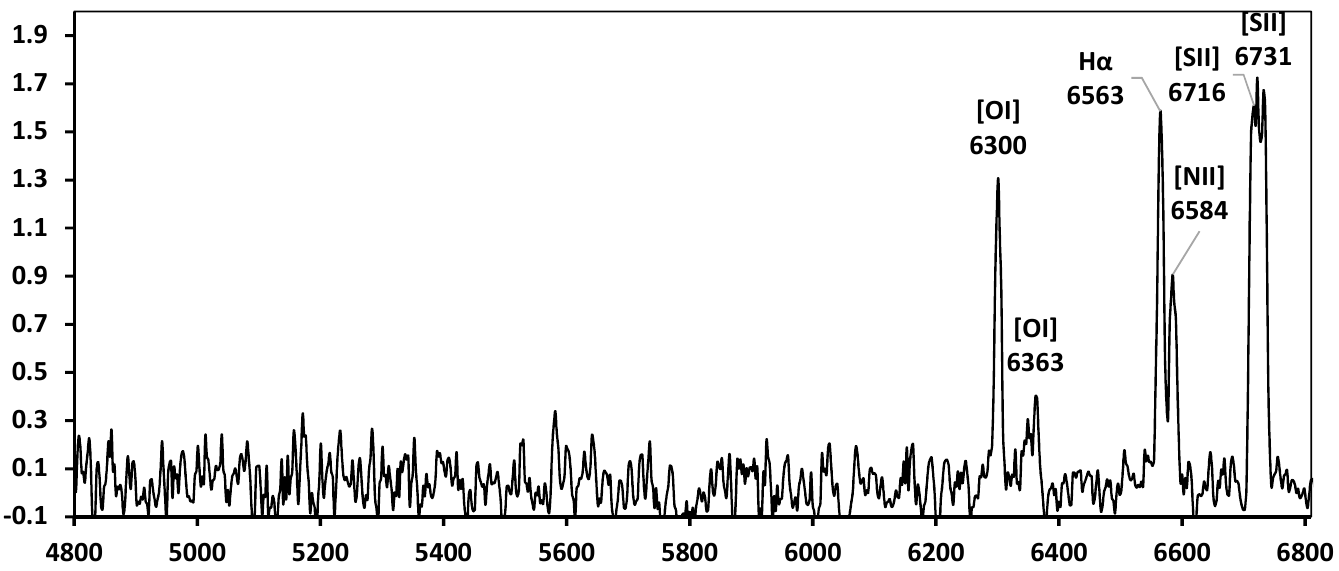}
\includegraphics[angle=0, width=8.5cm]{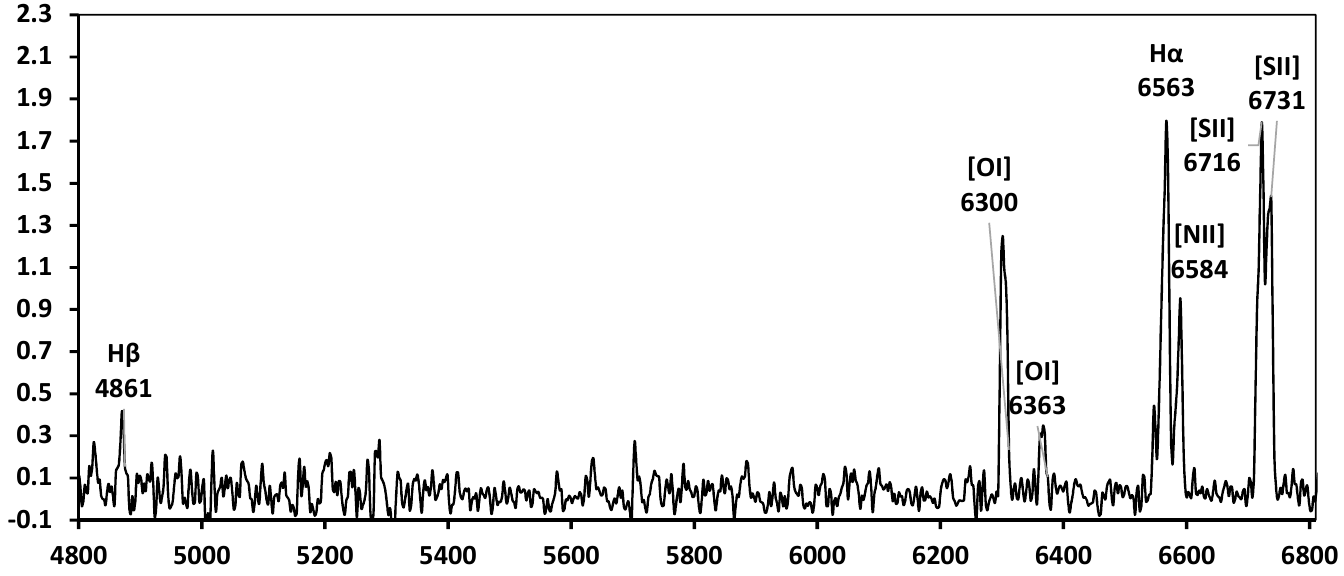}
\includegraphics[angle=0, width=8.5cm]{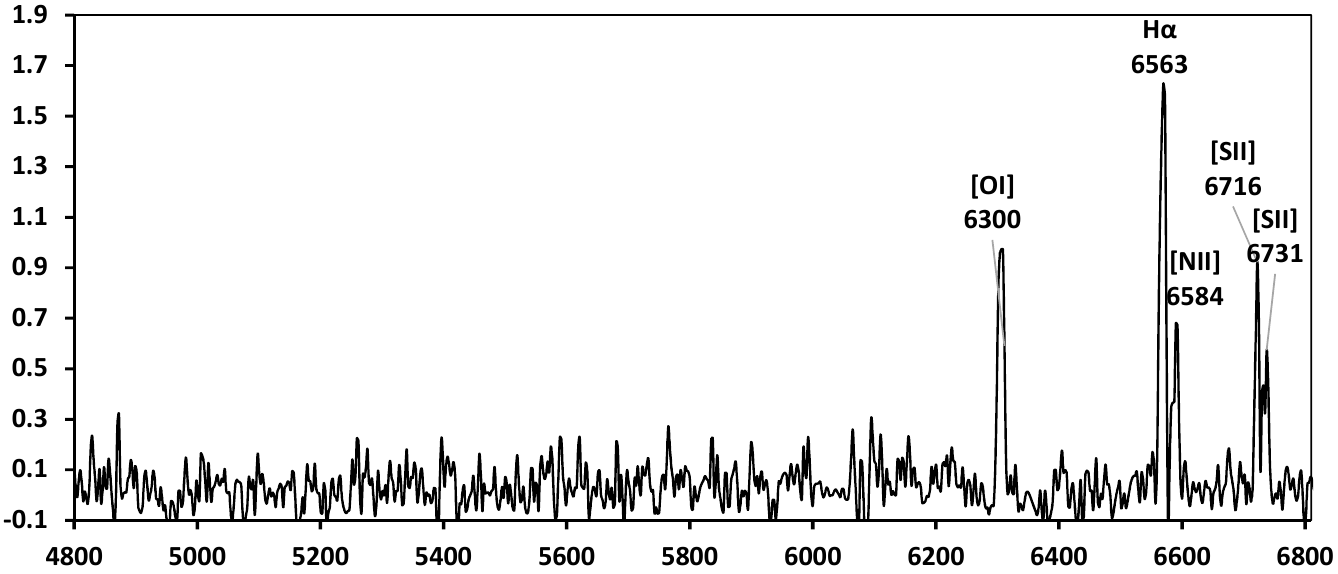}
\includegraphics[angle=0, width=8.5cm]{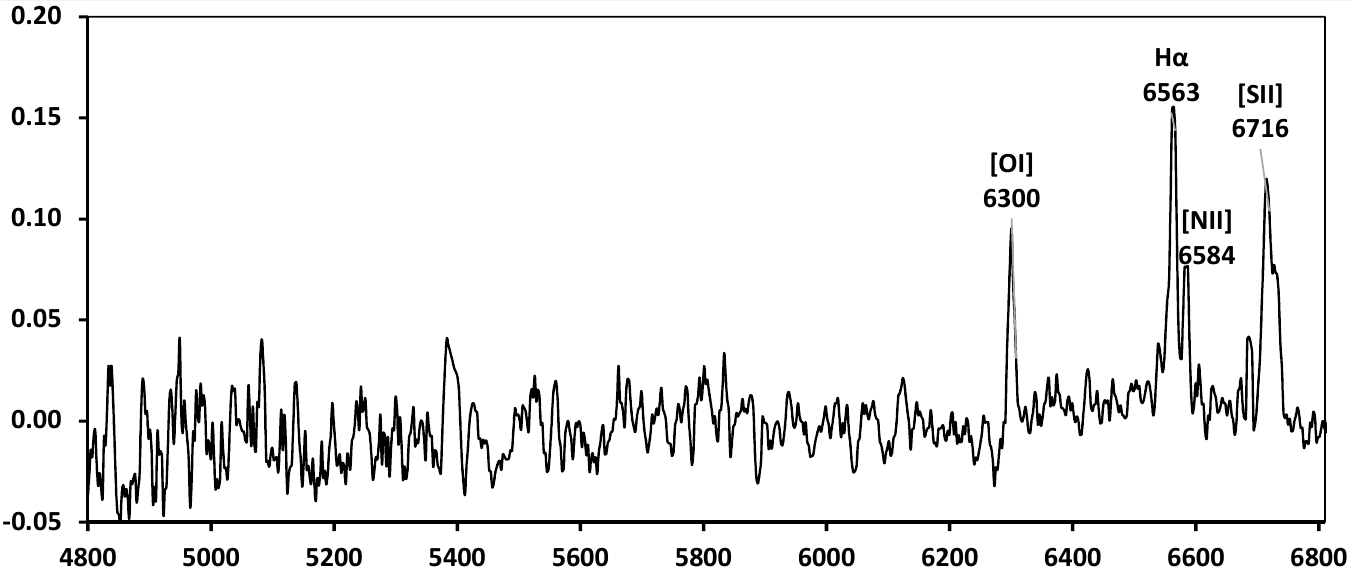}
\includegraphics[angle=0, width=8.5cm]{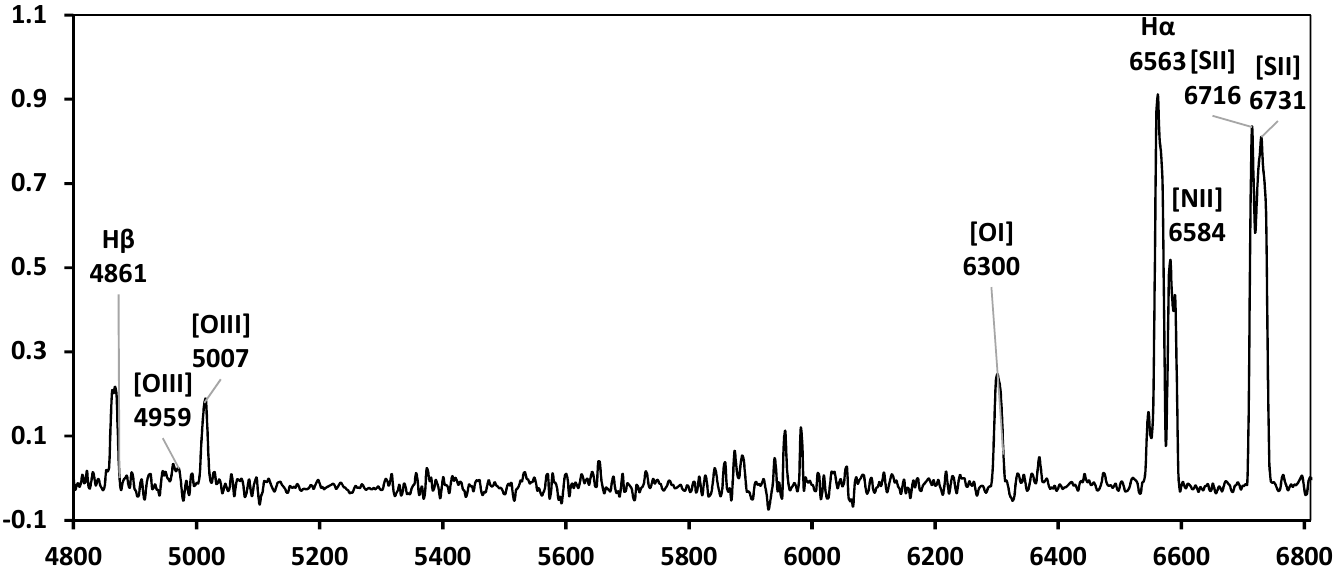}
\includegraphics[angle=0, width=8.5cm]{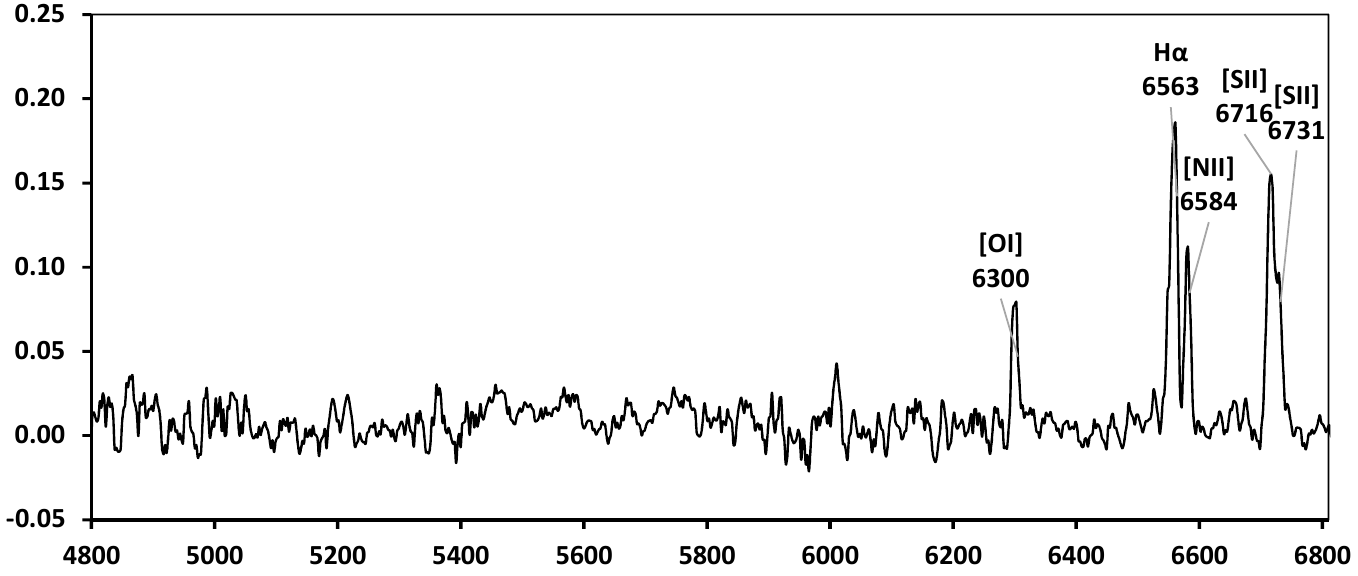}
\includegraphics[angle=0, width=8.5cm]{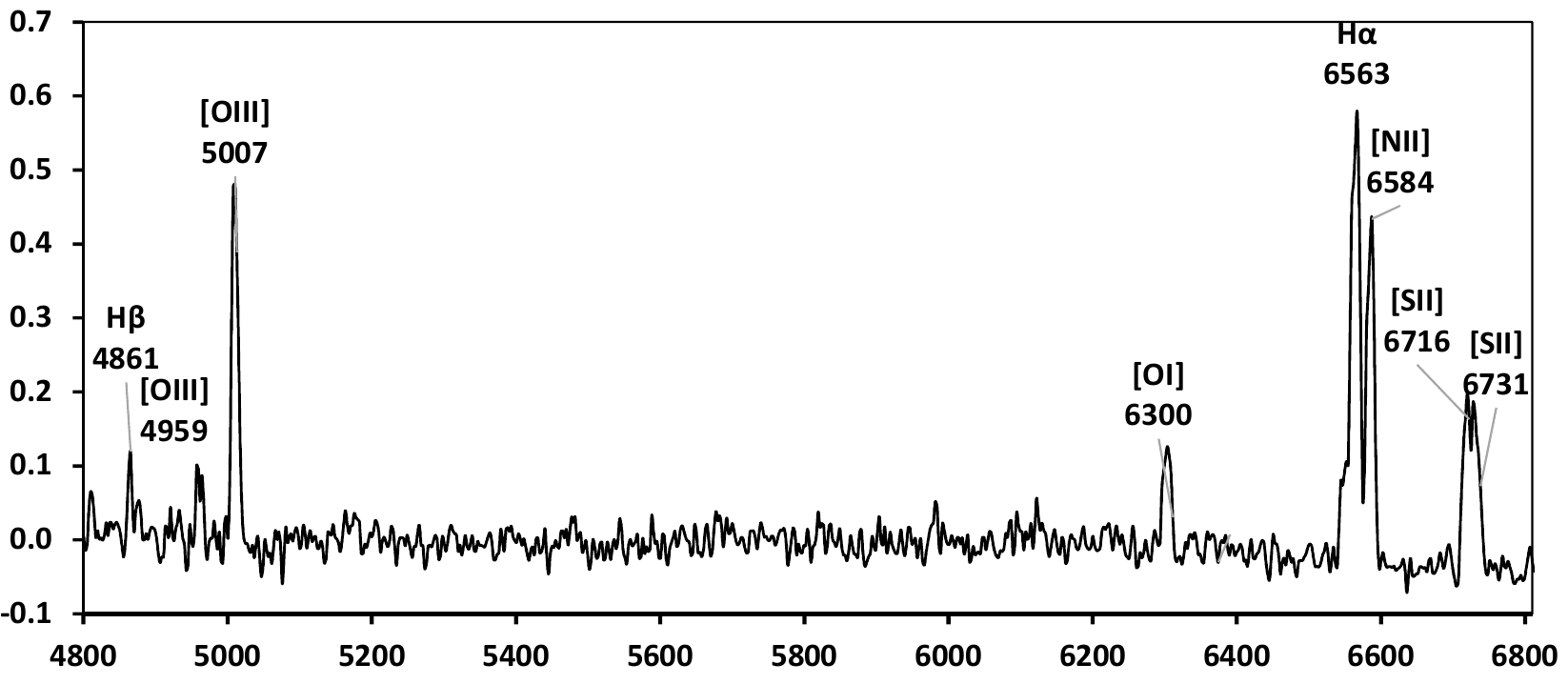}
\includegraphics[angle=0, width=8.5cm]{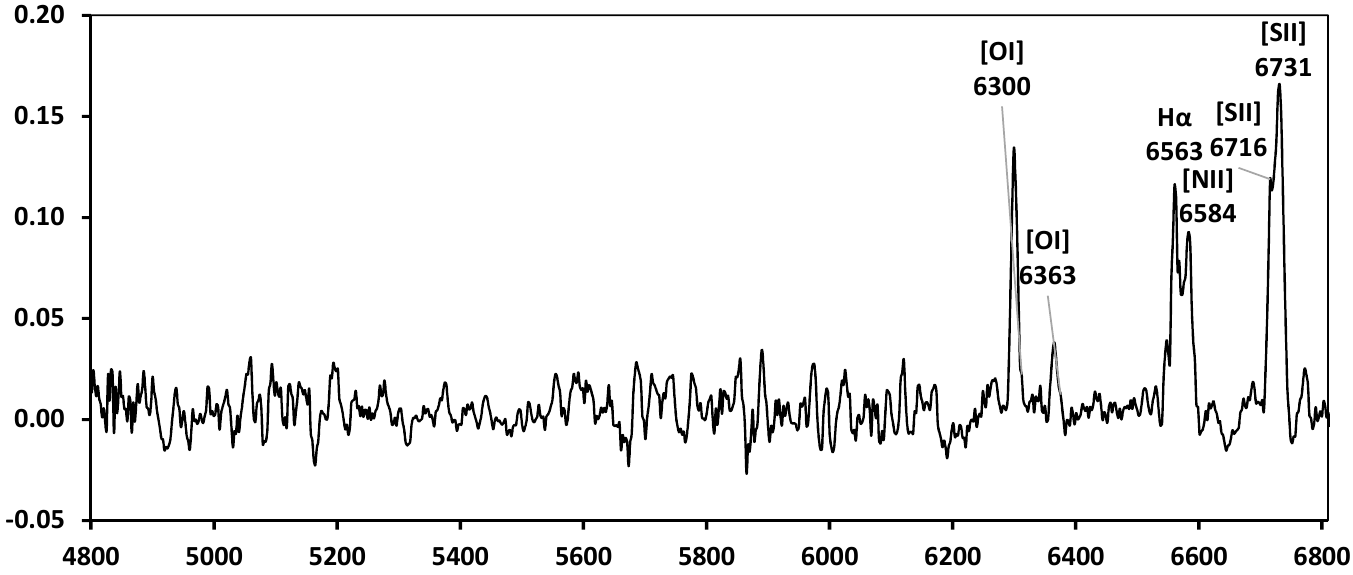}
\includegraphics[angle=0, width=8.5cm]{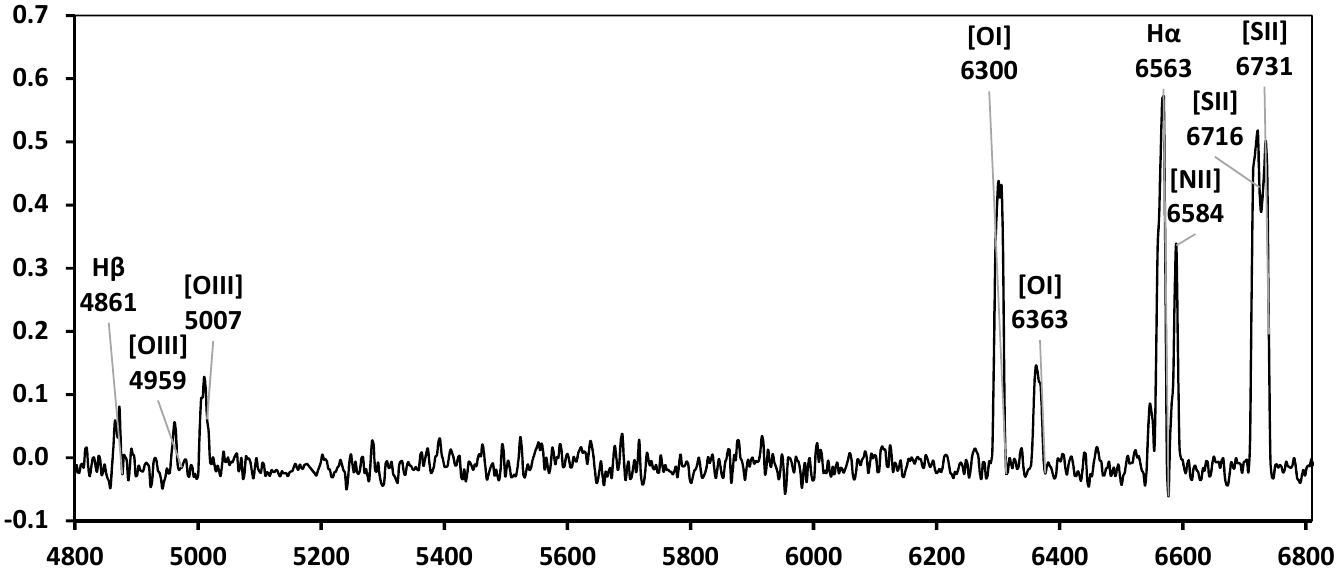}

\caption{Long-slit spectra of the NE rim taken with the RTT150 telescope. The flux (y axis) is in $10^{-15}$ ergs cm$^{-2}$ s$^{-1}$ \AA$^{-1}$. The wavelength (x-axis) is in Angstroms.}
\label{figure10}
\end{figure*}

\begin{figure*}
\includegraphics[angle=0, width=8.5cm]{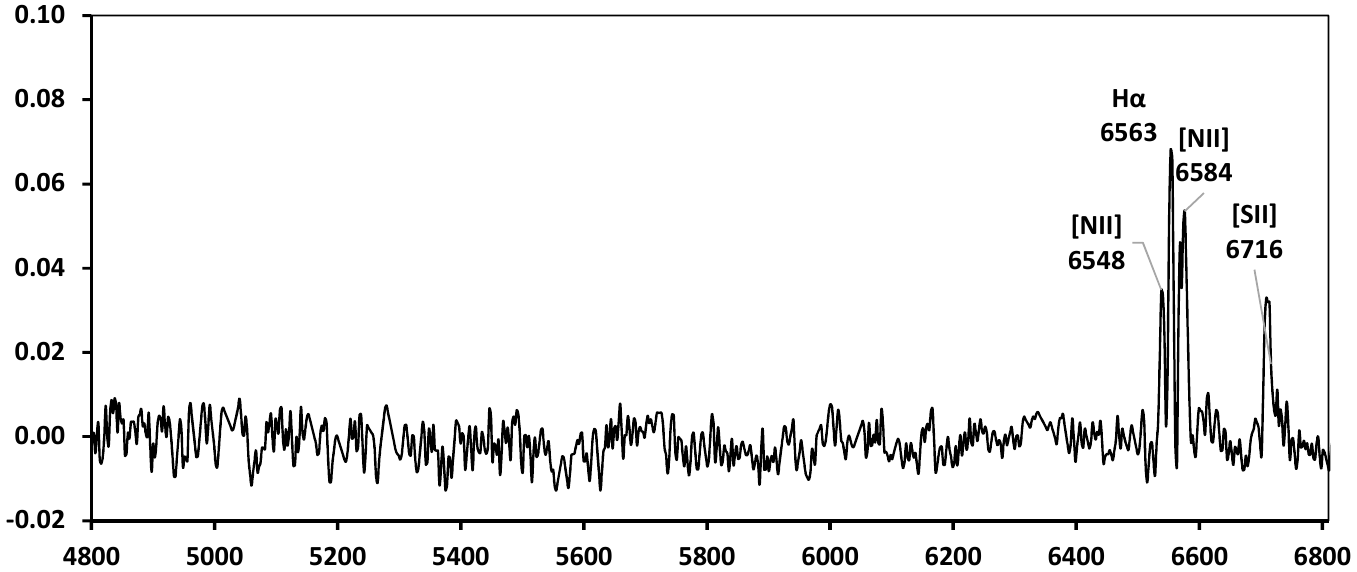}
\includegraphics[angle=0, width=8.5cm]{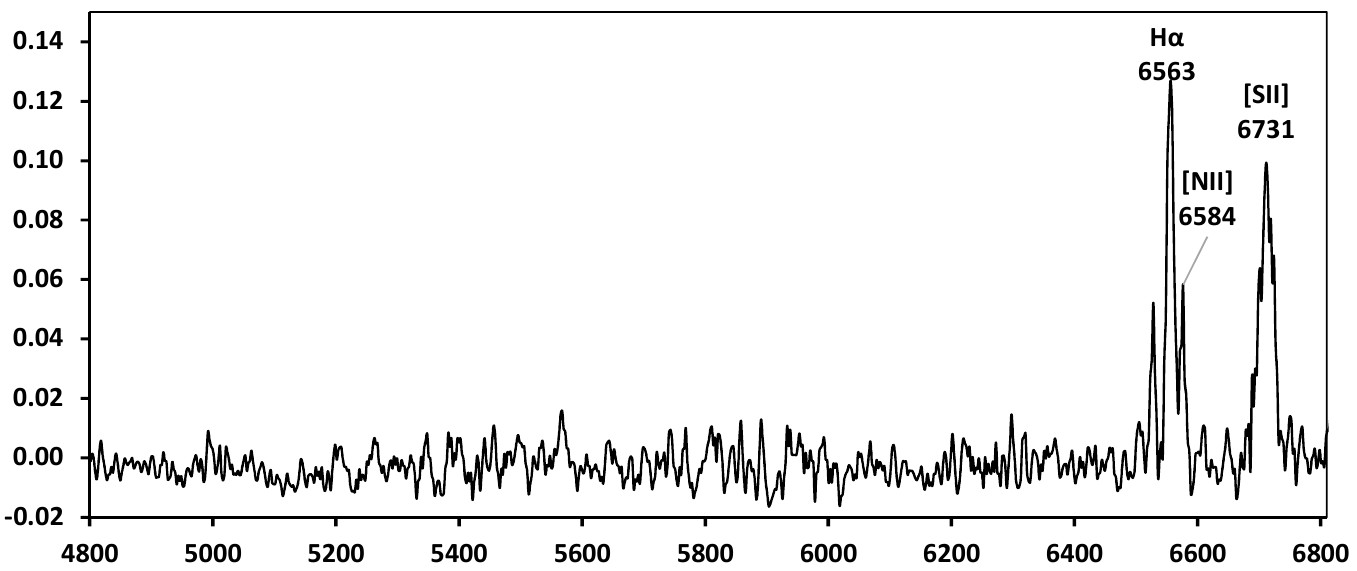}
\includegraphics[angle=0, width=8.5cm]{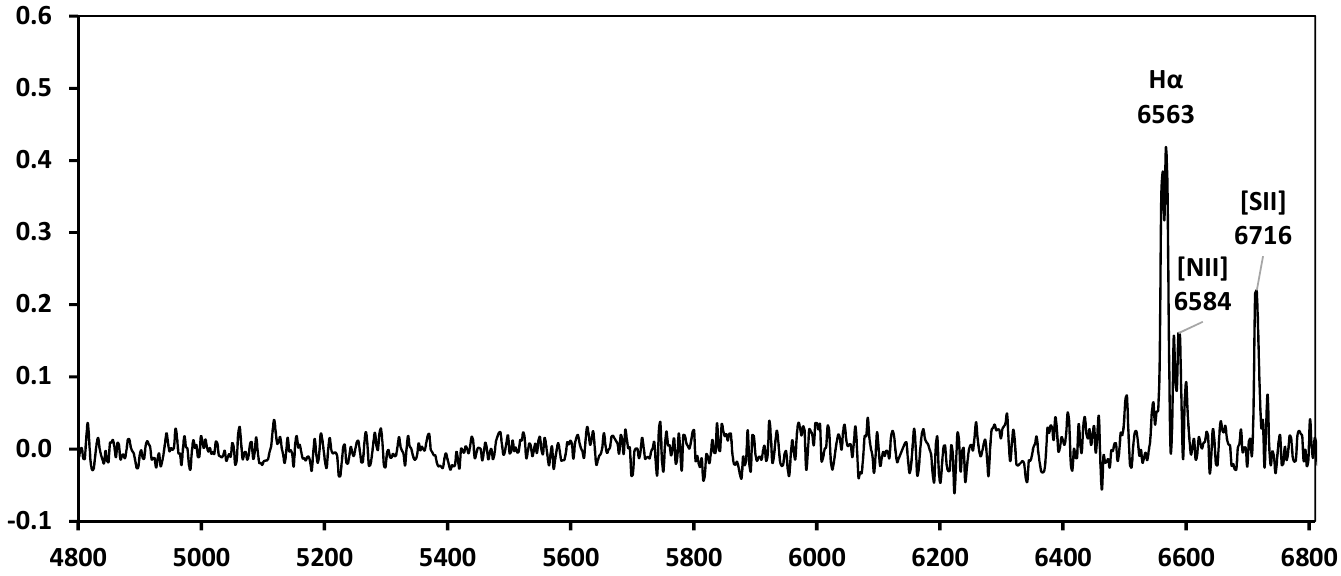}
\includegraphics[angle=0, width=8.5cm]{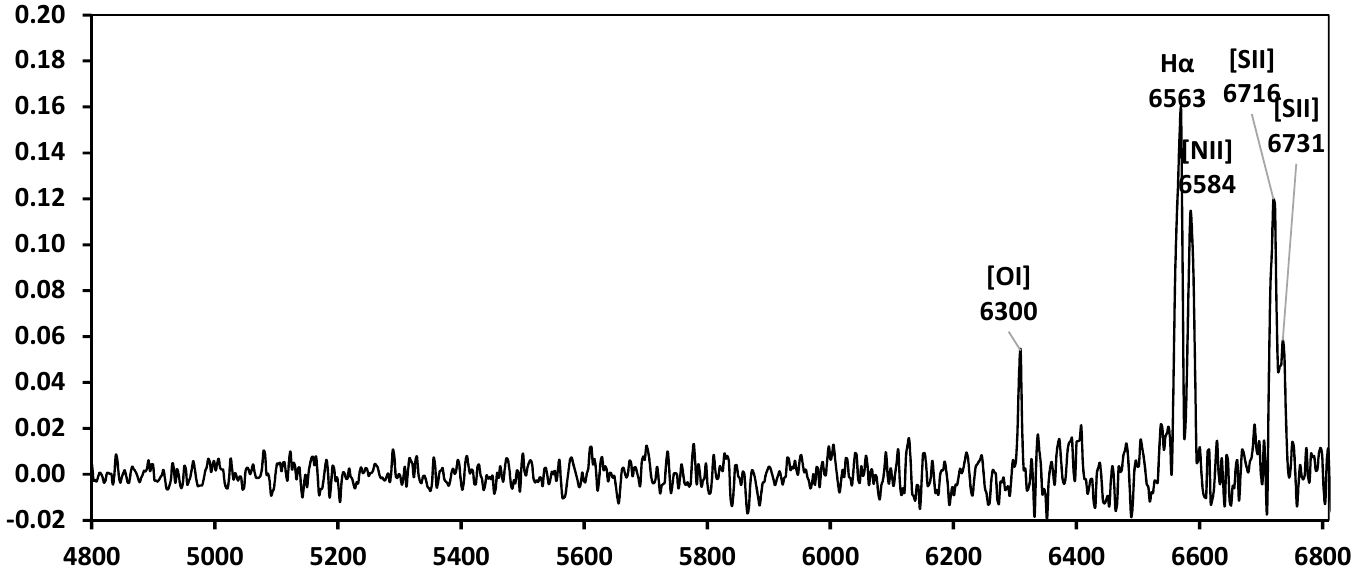}
\includegraphics[angle=0, width=8.5cm]{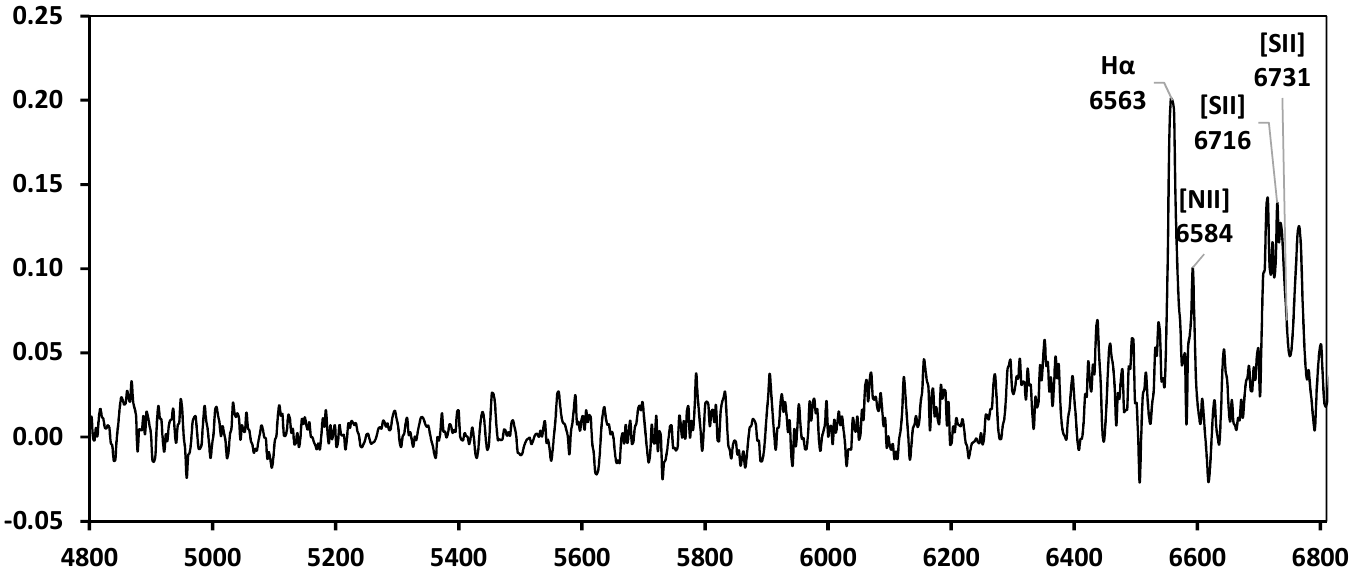}
\includegraphics[angle=0, width=8.5cm]{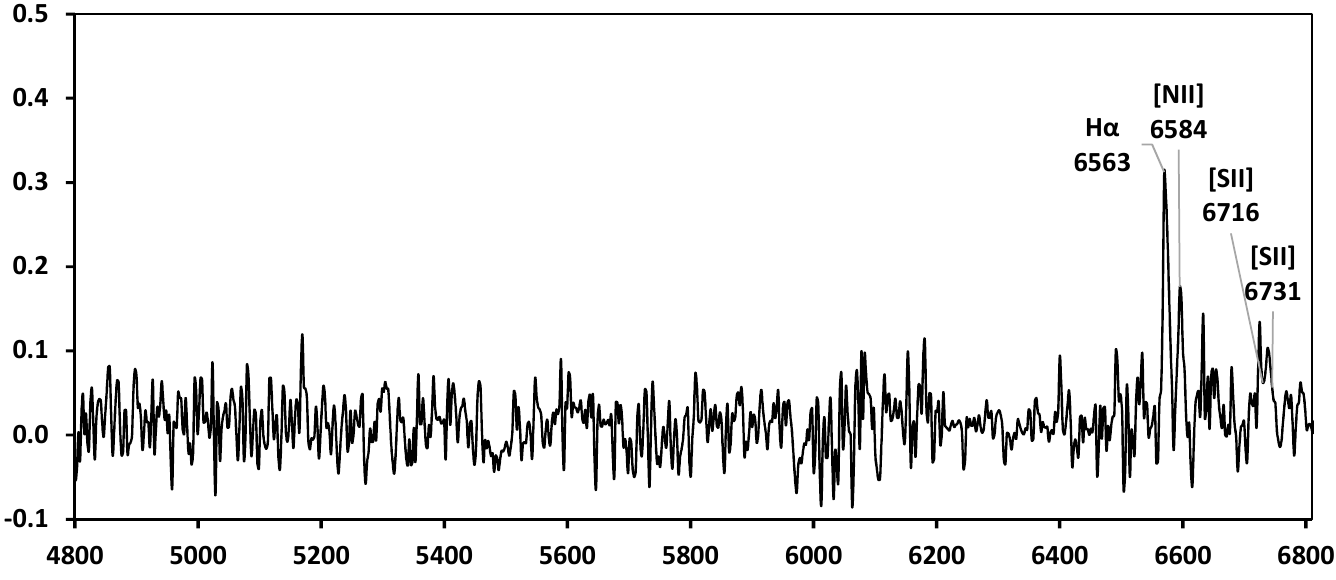}
\includegraphics[angle=0, width=8.5cm]{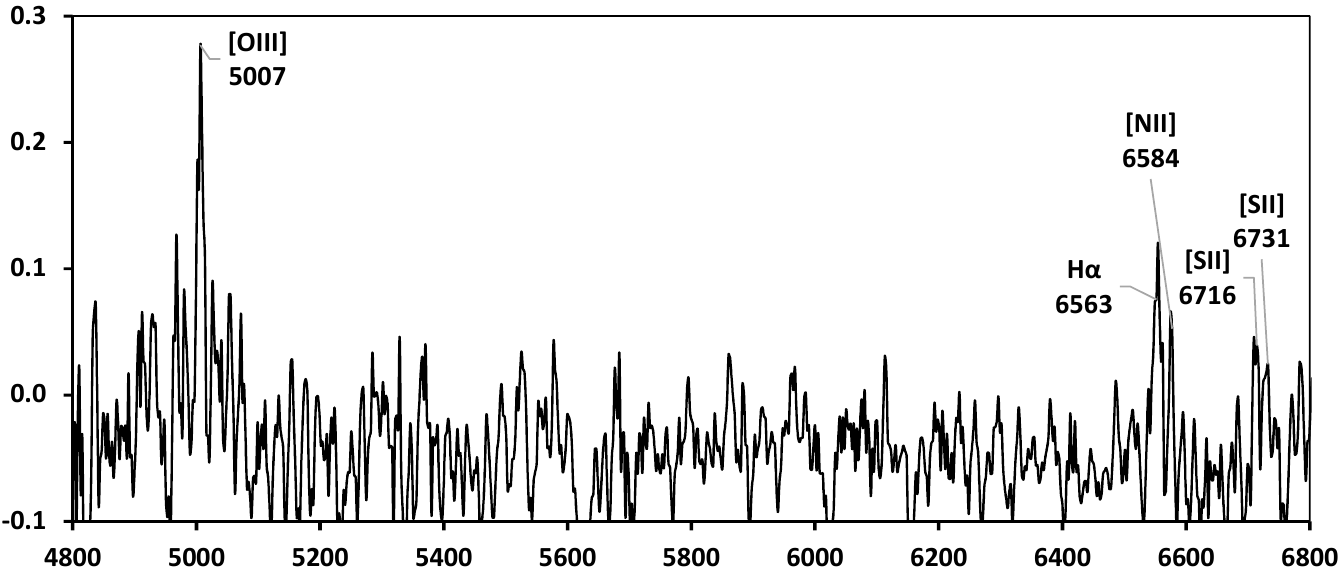}
\caption{Same as Fig. \ref{figure10}, for the NE filaments.}
\label{figure11}
\end{figure*}

\begin{figure*}
\includegraphics[angle=0, width=8.5cm]{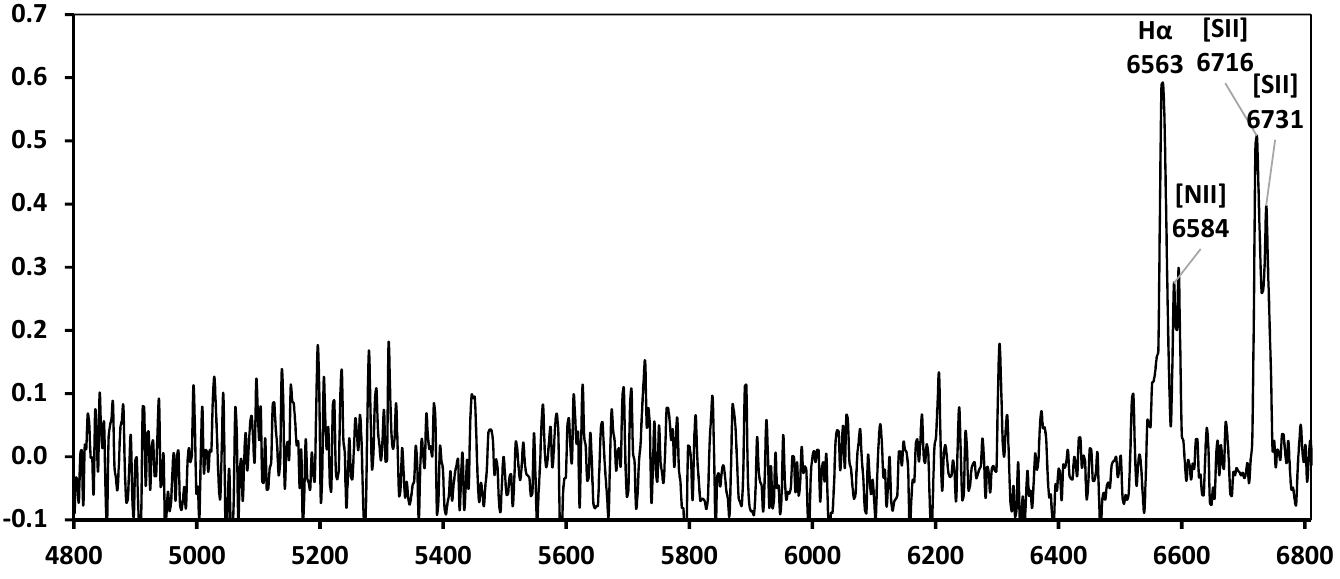}  
\includegraphics[angle=0, width=8.5cm]{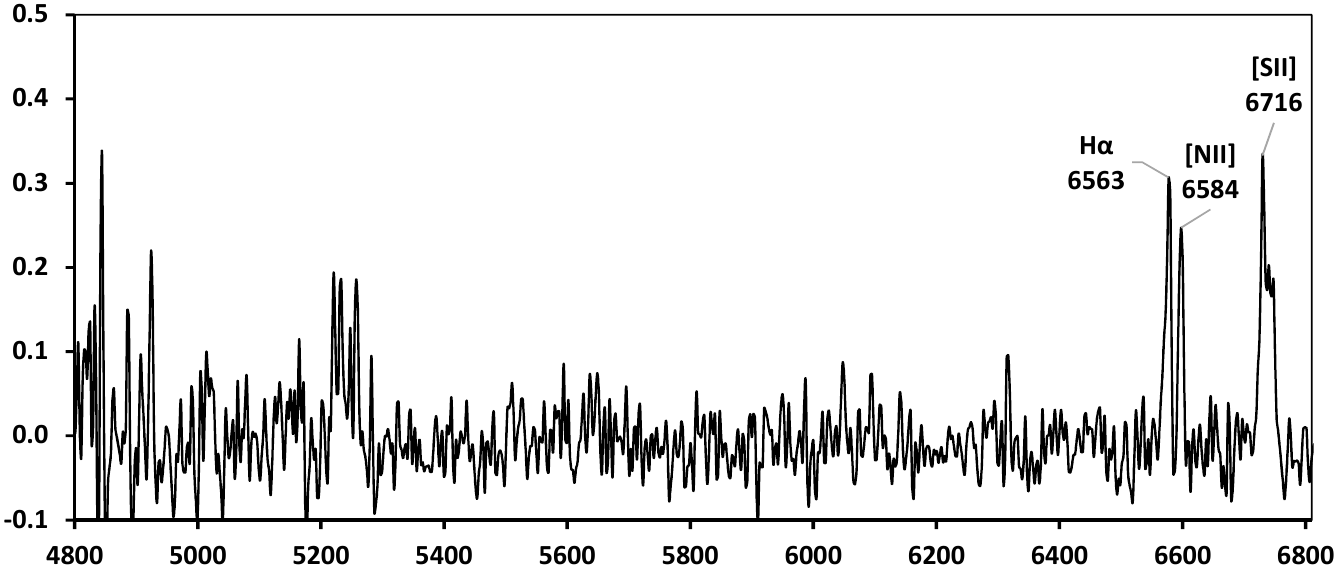}
\includegraphics[angle=0, width=8.5cm]{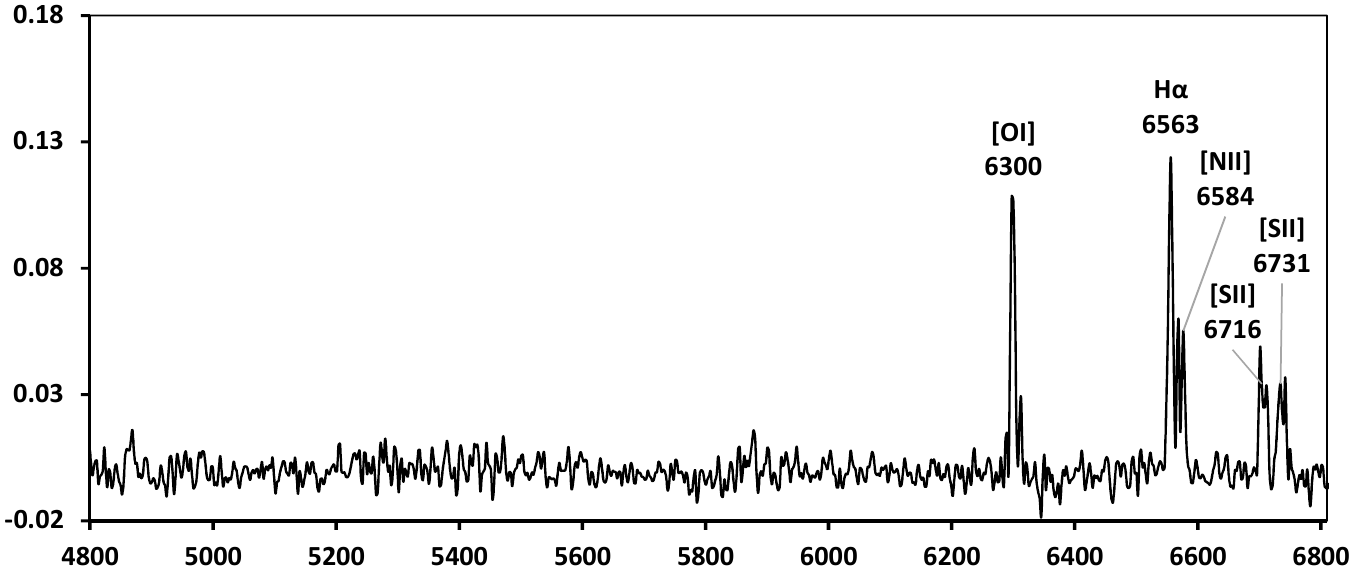}
\caption{Same as Fig. \ref{figure10}, for regions West and East.}
\label{figure12}
\end{figure*}

We obtained the long-slit spectra for the 22 spectral locations indicated on the images in Fig. \ref{figure8a}. We display the long-slit spectra for IC~443 and G189.6+3.3 in Figs \ref{figure10}$-$\ref{figure12}. The emission line fluxes with the line ratios are given in Tables \ref{Table3}$-$\ref{Table5}.

We present the spectra of the NE rim in Fig. \ref{figure10} (see Table \ref{Table3}). Our spectral observations regions correspond to Shell A in Figure 1 of \citet{Le08}.  We present the spectra of the relatively faint NE filaments in Fig. \ref{figure11} (see Table \ref{Table4}).  Our spectral observations regions correspond to the region in Figure 1 of \citet{Fe84}. The spectra for the West region of IC~443  and the East region of G189.6+3.3 are shown in Fig. \ref{figure12} (see Table \ref{Table5}). As can be seen in Figs \ref{figure10}$-$\ref{figure12}, there are no significantly different in the spectral features between the different regions.

\begin{table*}
\centering
\caption{Line fluxes relative to H$\alpha$ with the signal-to-noise ratios (S/N) for the radio bright NE rim. All line fluxes are normalized $F$(H$\alpha$) = 100. The [S\, {\sc ii}]/ H$\alpha$ ratio and electron density are given as well. The electron density $n_{\rm e}$ is derived from [S\,{\sc ii}]$\lambda$6716/$\lambda$6731 lines ratio.}
\label{Table3}
 \begin{tabular}{@{}p{3.5cm}p{1.8cm}p{1.8cm}p{1.8cm}p{1.8cm}p{1.8cm}p{1.8cm}@{}}
 \hline
 \hline
 	&	3i 	&	3j 	&	 	3l 	 & 3m	\\[0.5 ex]
\hline												
H$\beta$ ($\lambda$4861)	&	$...$	&	$...$	&			5 (10)	&	  $...$	 \\
												
$[$O$\,${\sc iii}$]$ ($\lambda$4959) 	&	$...$	&	$...$	&			$...$		&	  $...$	   \\
												
$[$O$\,${\sc iii}$]$ ($\lambda$5007) 	&	$...$	&	6 (20)	&			$...$	&	  $...$	   \\
												
$[$O$\,${\sc i}$]$ ($\lambda$6300) 	&	34 (20)	&	28 (40)	&			29 (52)	&	78 (19)		   \\
												
$[$O$\,${\sc i}$]$ ($\lambda$6363) 	&	 9 (14)	&	10 (20)	&		  11 (29)	&		41 (8)		    \\
												
$[$N$\,${\sc ii}$]$ ($\lambda$6548) 	&	 $...$	&	$...$	&		$...$	&	$...$		  \\
												
H$\alpha (\lambda$6563) 	&	100 (30)	&	100 (88)	&			100 (92)	&	100 (25)	    \\
												
$[$N$\,${\sc ii}$]$ ($\lambda$6584) 	&	29 (22)	&	19 (42)	&			18 (48)  &	52 (14)		  \\
												
$[$S$\,${\sc ii}$]$ ($\lambda$6716) 	&	41 (20)	&	27 (45)	&			79 (67)&	115 (24)	  \\
												
$[$S$\,${\sc ii}$]$ ($\lambda$6731) 	&	55 (24)	&	36 (45)	&	$...$		&	110 (22)		\\[0.5 ex]

F (H$\alpha$) ($10^{-15}$ erg~cm$^{-2}$~s$^{-1}$) &   2.83  & 12.20  &    23.0  &  21.0   \\
												
[S\, {\sc ii}]/ H$\alpha$ 	&	    0.96 $\pm$ 0.01	&	    0.63 $\pm$ 0.01	&	    		    0.79 $\pm$ 0.04 	&	    2.25 $\pm$ 0.02 	     \\[0.5 ex]
												
$n_{\rm e}$(cm$^{-3}$)	&	     1897 $\pm$ 50	&	    1859 $\pm$ 10	&	  	$...$	&	    573 $\pm$ 2		     \\
	\\																					
  \hline												
 \hline												
  	&	3p 	&	 	3s 	&			3b        & 3h              \\[0.5 ex] 
\hline												
												
H$\beta$ ($\lambda$4861)	&	12 (8)	&		$...$	&			 $...$   &   17 (38)    \\
												
$[$O$\,${\sc iii}$]$ ($\lambda$4959) 	&	$...$	&			$...$	&		 $...$    &  7 (14)   \\
												
$[$O$\,${\sc iii}$]$ ($\lambda$5007) 	&	$...$	&			$...$	&			 $...$    & 13 (33)    \\
												
$[$O$\,${\sc i}$]$ ($\lambda$6300) 	&	44 (24)	&			68 (18)	&			  47 (20)   &  22 (42)    \\
												
$[$O$\,${\sc i}$]$ ($\lambda$6363) 	&	10 (8)	&		$...$	&			 $...$  &      $...$    \\
												
$[$N$\,${\sc ii}$]$ ($\lambda$6548) 	&	$...$	&			$...$	&	     $...$    &   $...$  \\
												
H$\alpha (\lambda$6563) 	&	100 (33)	&			100 (30)	&	  100 (35)    & 100 (119)   \\
												
$[$N$\,${\sc ii}$]$ ($\lambda$6584) 	&	20 (17)	&			29 (9)	&		20 (25)   &  34 (72)     \\
												
$[$S$\,${\sc ii}$]$ ($\lambda$6716) 	&	73 (32)	&		38 (14)	&	 105 (25)    &    57 (95)   \\
												
$[$S$\,${\sc ii}$]$ ($\lambda$6731) 	&	52 (28)	&			26 (9)	&	$...$    &   72 (98)  \\[0.5 ex]

F (H$\alpha$) ($10^{-15}$erg cm$^{-2}$ s$^{-1}$) &  35.10  &    20.20 &   2.72   &    18.40  \\
												
[S\, {\sc ii}]/ H$\alpha$ 	&	    1.25 $\pm$ 0.01  	&	    		    0.64 $\pm$ 0.01  	&	  		    1.05 $\pm$ 0.01   &  1.29 $\pm$ 0.01    \\[0.5 ex]
												
$n_{\rm e}$(cm$^{-3}$)	&	    70 $\pm$ 2  	&			    22 $\pm$ 1  	&			 $...$   &  1546 $\pm$ 3  \\
\\[0.5 ex]											
  \hline
 \hline
  	&	3c 	&	 		3e 	&	3f 	&	3g  \\[0.5 ex]
\hline												
												
H$\beta$ ($\lambda$4861)	&	$...$	&			8 (24)	&	$...$	&	11 (18)    \\
												
$[$O$\,${\sc iii}$]$ ($\lambda$4959) 	&	$...$	&			8 (24)	&	$...$	&	6 (16)	 \\
												
$[$O$\,${\sc iii}$]$ ($\lambda$5007) 	&	$...$	&		23 (60)	&	$...$	&	18 (25) \\
												
$[$O$\,${\sc i}$]$ ($\lambda$6300) 	&	35 (15)	&			14 (30)	&	63 (24)	&	70 (65)  \\
												
$[$O$\,${\sc i}$]$ ($\lambda$6363) 	&	$...$	&			$...$	&	20 (15)	&	24 (29)	 \\
												
$[$N$\,${\sc ii}$]$ ($\lambda$6548) 	&	$...$	&		$...$	&	$...$	&	$...$	 \\
												
H$\alpha (\lambda$6563) 	&	100 (35)	&		100 (68)	&	100 (25)	&	100 (82)	 \\
												
$[$N$\,${\sc ii}$]$ ($\lambda$6584) 	&	34 (24)	&			31(55)	&	48 (2)	&	24 (44)	 \\
												
$[$S$\,${\sc ii}$]$ ($\lambda$6716) 	&	59 (30)	&		18 (30)	&	55 (24)	&	76 (72)	 \\
												
$[$S$\,${\sc ii}$]$ ($\lambda$6731) 	&	66 (22)	&			22 (22)	&	92 (31)	&	69 (60)	 \\[0.5 ex]

F (H$\alpha$) ($10^{-15}$erg cm$^{-2}$ s$^{-1}$) &   3.10 &  14.80 & 2.87 &  9.69  \\
												
[S\, {\sc ii}]/ H$\alpha$ 	&	    1.25 $\pm$ 0.02	&	      0.40 $\pm$ 0.04	&	    1.47 $\pm$ 0.09	&	    1.45 $\pm$ 0.01		     \\[0.5 ex]
												
$n_{\rm e}$(cm$^{-3}$)	&	    1023 $\pm$ 5	&			    1384 $\pm$ 8	&	    4384 $\pm$ 299	&	    458 $\pm$ 2		     \\											
  \hline	
\hline
\end{tabular}
\end{table*}

\begin{table*}
\centering
\caption{Same as Table \ref{Table3}, for the NE filaments.}
\label{Table4}
 \begin{tabular}{@{}p{3.5cm}p{1.6cm}p{1.6cm}p{1.6cm}p{1.6cm}p{1.6cm}p{1.6cm}p{1.6cm}@{}} 
 \hline	
 \hline	
  		            &	7d 	&	7c 	&	7e 	&	7f 	&	7g 	 & 7a	&	7b \\[0.5 ex]
                 \hline																						
H$\beta$ ($\lambda$4861)		        &	$...$	&	$...$	&	$...$	&	$...$	&	$...$	 &	 $...$	&	 $...$	   \\													
$[$O$\	${\sc iii}$]$ ($\lambda$4959) 	&	$...$	&	$...$	&	$...$	&	$...$	&	$...$	 &	  $...$	&	  $...$	   \\												
$[$O$\	${\sc iii}$]$ ($\lambda$5007) 	&	$...$	&	$...$	&	$...$	&	$...$	&	$...$	 &	  $...$	&	183 (10)	   \\													
$[$O$\	${\sc i}$]$ ($\lambda$6300) 	&	$...$	&	$...$	&	$...$	&	17 (13)	&	$...$	 &	  $...$	&	  $...$	   \\
													
$[$O$\	${\sc i}$]$ ($\lambda$6363) 	&	$...$	&	$...$	&	$...$	&	$...$	&	$...$	 &	 $...$	&	 $...$	   \\
													
$[$N$\	${\sc ii}$]$ ($\lambda$6548) 	&	62 (16) 	&	$...$	&	$...$	&	$...$	&	$...$	 &	  $...$	&	  $...$	    \\													
H$\alpha (\lambda$6563) 		&	100 (26)	&	100 (26)	&	100 (22)	&	100 (24)	&	100 (19)	&	100 (20)	&	100 (12)	    \\
													
$[$N$\	${\sc ii}$]$ ($\lambda$6584) 	&	133 (23)	&	33 (14)	&	34 (11)	&	45 (18)	&	23 (10)		&	51 (10)	&	36 (11)	   \\	
													
$[$S$\	${\sc ii}$]$ ($\lambda$6716) 	&	114 (14)	&	$...$	&	50 (13)	&	50 (19)	&	53 (12) &	58 (10)	&	75 (10)			  \\													
$[$S$\	${\sc ii}$]$ ($\lambda$6731) 	&	$...$	&	139 (21)	&	$...$	&	35 (12) &	73 (11)	&	48 (9)	&	55 (9)	  \\[0.5 ex]

F (H$\alpha$) ($10^{-15}$erg cm$^{-2}$ s$^{-1}$) &   0.63  & 1.90 &  5.80 &  2.09 & 3.52 &  3.44 &  3.05\\
													
[S\	 {\sc ii}]/ H$\alpha$ 	&	    1.14 $\pm$ 0.01	&	    1.39 $\pm$ 0.01	&	    0.50 $\pm$ 0.01	&	 0.85 $\pm$ 0.01		&	    1.26 $\pm$ 0.01 &	    1.06 $\pm$ 0.08  	&	    1.30 $\pm$ 0.03 	      \\[0.5 ex]	
													
$n_{\rm e}$(cm$^{-3}$)     	 	&	$...$	&	   $...$	&	$...$	&	49 $\pm$ 1	&	    2078 $\pm$ 9 &	    286 $\pm$ 17  	&	    107 $\pm$ 2        \\										
\hline
 \hline
\end{tabular}
\end{table*}

\begin{table*}
\centering
\caption{Same as Table \ref{Table3}, for regions West and East.}
\label{Table5}
 \begin{tabular}{@{}p{3.5cm}p{1.8cm}p{1.8cm}p{1.8cm}p{1.8cm}p{1.8cm}@{}}
  
  \hline
 \hline
 					&	4a	&	4b	&	 11a	               \\[0.5 ex]

  					&	West	&	West	&	East	\\[0.5 ex]
\hline	
											
H$\beta$ ($\lambda$4861)		&		  $...$	&	 $...$			&	 $...$	 \\
												
$[$O$\,${\sc iii}$]$ ($\lambda$4959) 	&	  	$...$	&	 $...$			&	 $...$  \\
												
$[$O$\,${\sc iii}$]$ ($\lambda$5007) 	&		  $...$	&	 $...$			&	 $...$		   \\
												
$[$O$\,${\sc i}$]$ ($\lambda$6300) 	&		 $...$	&	  $...$			&	14 (22)      \\
												
$[$O$\,${\sc i}$]$ ($\lambda$6363) 	&		  $...$	&	  $...$			&	 $...$     \\
												
$[$N$\,${\sc ii}$]$ ($\lambda$6548) 	&		 $...$	&	 $...$			&	$...$	    \\
												
H$\alpha (\lambda$6563) 		&		100 (15)&	100 (15) 	  	&	100 (24)    \\
												
$[$N$\,${\sc ii}$]$ ($\lambda$6584) 	&		17 (7)	&	40 (12)			&	32 (17)	     \\
												
$[$S$\,${\sc ii}$]$ ($\lambda$6716) 	&		36 (13)	&	$...$			&	42 (11)	  \\
												
$[$S$\,${\sc ii}$]$ ($\lambda$6731) 	&		40 (10)	&	151 (15)		&	44 (10)	  \\[0.5 ex]

F (H$\alpha$) ($10^{-15}$ erg cm$^{-2}$ s$^{-1}$) &   15.60  & 4.64 &  1.19 \\
												
[S\, {\sc ii}]/ H$\alpha$ 		&	    0.76 $\pm$ 0.01 	& 1.51 $\pm$ 0.02 	&	 0.86 $\pm$ 0.01	 \\[0.5 ex]
												
$n_{\rm e}$(cm$^{-3}$)		&	     1000 $\pm$ 5  	& $...$			&	   811 $\pm$ 2	    \\									
  \hline												
 \hline			
\end{tabular}
\end{table*}

%\vspace{10mm}

\subsection{H\,{\sc i} analysis and results}
\begin{figure*}
\includegraphics[angle=0, width=12.5cm]{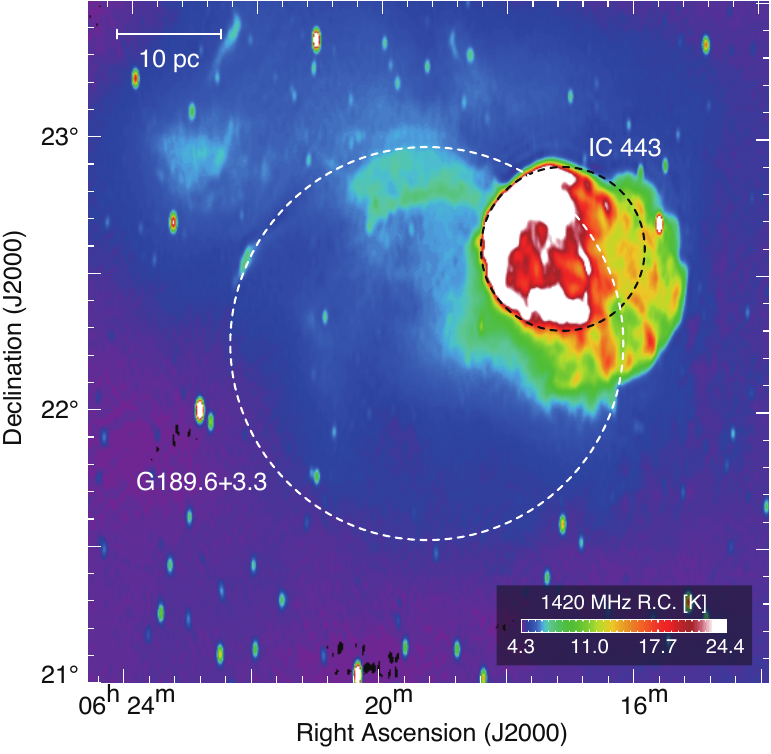}
\caption{Distribution of the 1.4~GHz radio continuum from the CGPS \citep{Ta03}. The two dashed circles indicate radio shell boundaries of the SNRs G189.6$+$3.3 and IC~443, which were determined by eye inspection. The scale bar shows 10 pc in length assuming the source distance of 1.3 kpc.}
\label{figure13}
\end{figure*}

Before the H\,{\sc i} analysis, we determined the radio-shell boundaries of IC~443 and G189.6$+$3.3. Fig.~\ref{figure13} shows the 1.4 GHz radio continuum toward the IC~443 and G189.6$+$3.3 regions. We fitted the radio shell boundaries by eye inspection in order to compare them with the H\,{\sc i} distributions. This process gives the center position and radius of the radio shell as ($\alpha_\mathrm{J2000}$, $\delta_\mathrm{J2000}) \sim (06^\mathrm{h}17^\mathrm{m}07\fs04$, $+22^{\circ}36'07\farcs9$) and $\sim$0.30 degree for IC~443; ($\alpha_\mathrm{J2000}$, $\delta_\mathrm{J2000}) \sim (06^\mathrm{h}19^\mathrm{m}16\fs20$, $+22^{\circ}15'14\farcs9$) and $\sim$0.72 degrees for G189.6$+$3.3, respectively.

\begin{figure*}
\includegraphics[angle=0, width=18cm]{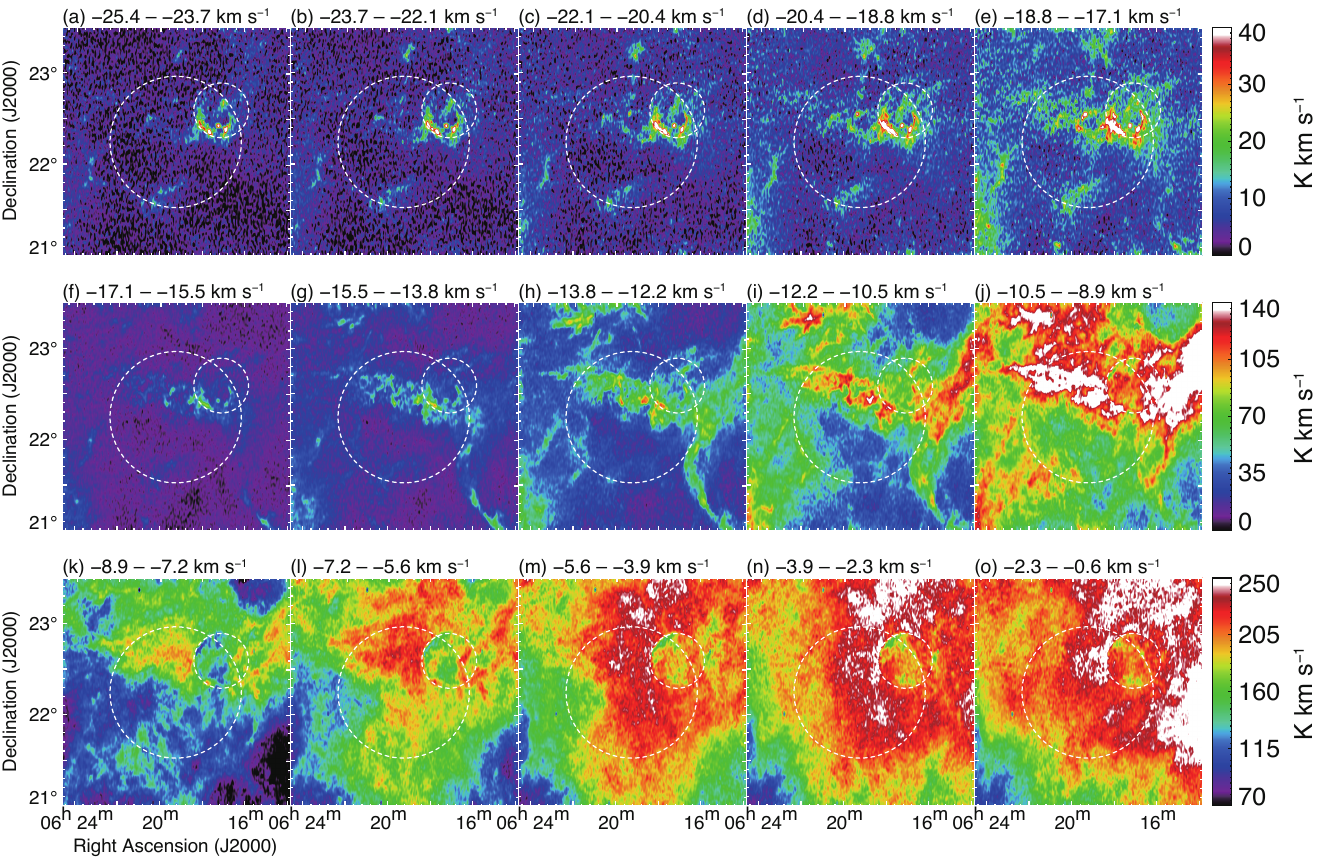}
\caption{Velocity channel distributions of H\,{\sc i} superposed on the radio shell boundaries of G189.6$+$3.3 and IC~443. Each panel shows H\,{\sc i} integrated intensity every 1.7~km~s$^{-1}$ in a velocity range from $-25.4$ to $-0.6$~km~s$^{-1}$. The intensity color scales are different for each row.}
\label{figure14}
\end{figure*}

Fig.~\ref{figure14} shows the velocity channel maps of H\,{\sc i} superposed on the radio continuum contours of IC~443 and G189.6$+$3.3. We confirmed that H\,{\sc i} absorption features due to the strong radio continuum of IC~443 are seen at $V_\mathrm{LSR}~\gtrsim~-10$~km~s$^{-1}$ (see Figs~\ref{figure14}(j)--\ref{figure14}(o)). We can see shocked H\,{\sc i} clouds of IC~443 at $V_\mathrm{LSR}~\lesssim~-15$~km~s$^{-1}$ (see Figs~\ref{figure14}(a)--\ref{figure14}(f)), which is consistent with the previous studies (e.g. \citealt{Le08}). We also newly found some shell-like distributions of H\,{\sc i} that are possibly associated with G189.6$+$3.3: the southeastern shell in Fig.~\ref{figure14}(j) and the western shell in Figs~\ref{figure14}(g) and \ref{figure14}(h). These clumpy and/or sharp edges of H\,{\sc i} seem to be along the boundary of the radio-continuum shell.

\begin{figure*}
\includegraphics[angle=0, width=12.5cm]{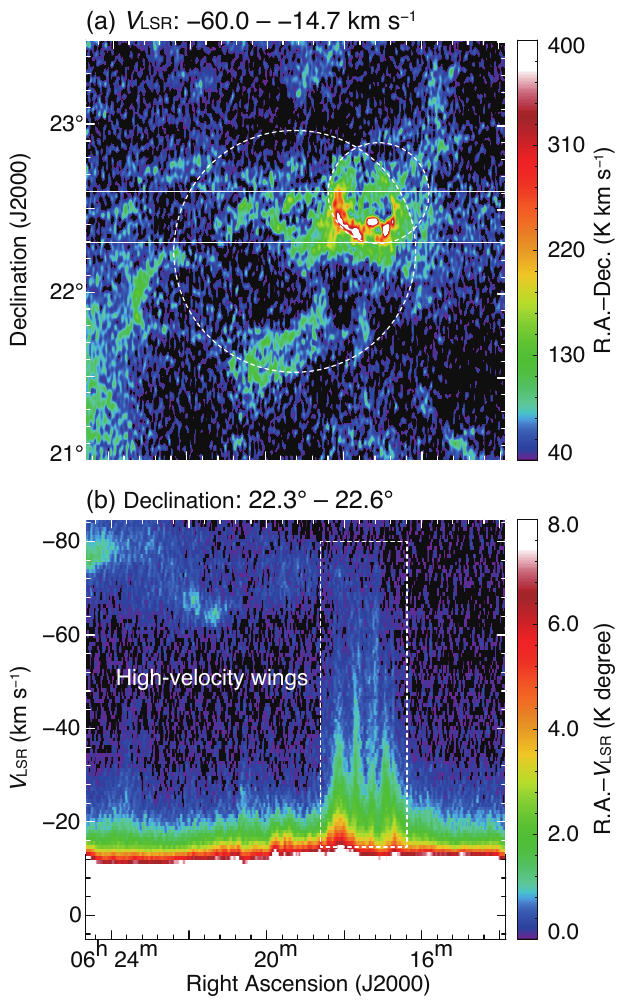}
\caption{(a) Velocity integrated intensity map of H\,{\sc i} superposed on the radio shell boundaries of G189.6$+$3.3 and IC~443. The integration velocity range of H\,{\sc i} is from $-60$ to $-14.7$ km s$^{-1}$. (b) Position--velocity diagram of H\,{\sc i}. The integration Declination range is from 22\fdg3 to 22\fdg6. The region enclosed by the dashed box indicates high-velocity wings of H\,{\sc i} (see the text).}
\label{figurepv}
\end{figure*}

Fig.~\ref{figurepv}(a) shows the velocity integrated map of shocked H\,{\sc i} clouds. We confirmed that the shocked clouds are mainly distributed toward the overlap region of the two SNRs. Fig.~\ref{figurepv}(b) shows the position--velocity diagram of H\,{\sc i}. When integrating over the spatial range of the shocked H\,{\sc i} clouds, we can reveal broad H\,{\sc i} components in the position--velocity diagram, whose velocity range is from $\sim$$-20$ to $\sim$$-80$~km~s$^{-1}$. These broad components are generally known to be high-velocity wings due to the gas acceleration by supernova shocks \citep[e.g.][]{1991ApJ...382..204K}.

\clearpage
\section{Discussion and Conclusions}
\label{discuss}
We presented optical photometric and spectroscopic observations of IC~443 and G189.6+3.3 regions, covering 2020$-$2023 observations. We also analyzed the archival H\,{\sc i} data to investigate the atomic environment of these SNRs. Motivated by these observations, we have constructed their properties and the local ambient medium. 

\subsection{H$\alpha$ morphology}
We investigated H$\alpha$ emission in eight regions (Regions 1, 2, 5, 7, 8, 9, 10, and 11) around the G189.6+3.3 and in three regions (Regions 3, 4, and 6) around the IC~443 indicated in Fig. \ref{radio}. We should note that G189.6+3.3, in particular in Region 3, overlaps with IC~443. The H$\alpha$ images of IC~443 and G189.6+3.3 show diffuse and filamentary structures. Many of the bright H$\alpha$ filaments are visible in Figs \ref{figure1}$-$\ref{figure8}. We also observed the {\it Suzaku} FoV region (shown by a magenta box in Fig. \ref{radio}), but we did not detect any optical emission from the region with the 900-s exposure times. 

\subsection{Optical spectroscopy and emission line ratios}
Our optical spectroscopic study covers the NE rim, NE filaments, and other two regions (the West region of IC~443 and the East region of G189.6+3.3) showing strong emission from lines typically observed in SNRs, such as H$\beta$$\lambda$4861, [O\,{\sc iii}]$\lambda$4959, $\lambda$5007, [O\,{\sc i}]$\lambda$6300, $\lambda$6363, H$\alpha$$\lambda$6563, [N\,{\sc ii}]$\lambda$6584 and [S\,{\sc ii}]$\lambda$6716, $\lambda$6731 (see Figs \ref{figure10}$-$\ref{figure12}). 

\textbf{\textit{NE rim}}. Optical spectra of the NE rim exhibit [S\,{\sc ii}]/H$\alpha$ ratios in range 0.4$-$2.9, confirming emission from shock-heated gas (e.g. \citealt{Ra79, Fe85, BlLo97}). The electron density ($n_{\rm e}$) can be obtained from the [S\,{\sc ii}] 6716/6731 flux ratio \citep{OsFe06}. Using the \texttt{temden} routine and the average of the [S\,{\sc ii}] 6716/6731 flux ratio, we found electron density for the regions as $n_{\rm e}$ $\sim$ 20$-$1900 cm$^{-3}$ (for electron temperature $T$=$10^{4}$ K), except for the slit position 3f indicating a very dense medium ($n_{\rm e}$ $\sim$ 4384 cm$^{-3}$). These values are representative of intrinsic fluctuations in the density with different values obtained for different regions. Our electron density range is in agreement with previous electron density estimates (i.e. $\sim$100$-$500 cm$^{-3}$: \citealt{Fe80}, and $\sim$100$-$2500 cm$^{-3}$: \citealt{AlDr19}). 

Based on a planar shock model of \citet{Ha87} and using the flux of $[$O$\,${\sc iii}$]$ $\lambda$5007 line relative to  H$\beta$, we estimate the shock velocity $V_{\rm s}$ $\sim$ 80~km~s$^{-1}$ for different filaments (3e, 3g and 3h). In the following, we found the pre-shock cloud density ($n_{\rm c}$) of $\sim$16$-$54 cm$^{-3}$ using Equation (7) in \citet{Fe80}. Our pre-shock cloud density values are consistent with the results reported by \citet{AlDr19}, but higher than those estimated by \citet{Fe80}. Assuming that $E(B-V)=0.664c$, where $c$ is a logarithmic extinction \citep{Ka76, Al84},  we found extinction $E(B-V)$ value of $\sim$0.89$-$1.65 for the NE rim (slit positions 3p, 3e, 3g and 3l). These values are in good agreement with extinction values found by \citet{Fe80} and \citet{AlDr19}. From the relation of \citet{Pr95}, we derived the column density $N_{\rm H}$ of $\sim$ (4.8$-$8.9)$\times$ $10^{21}$ cm$^{-2}$. \\

\textbf{\textit{NE filaments}}. For the NE filaments, we estimated [S\,{\sc ii}]/H$\alpha$ ratios in the range of 0.5$-$1.4, which confirms strong shock-heated emission (see \citealt{Fe84}). We obtained electron density for slit positions 7a, 7b, 7g, and 7f  as $n_{\rm e}$ $\sim$ 49$-$2078~cm$^{-3}$ (see Table \ref{Table4}). The presence strong [O\,{\sc iii}] $\lambda$5007 emission in spectra at 7b suggests high shock velocity.  In contrast, spectra at other slit locations showed no appreciable [O\,{\sc iii}] emission, indicating much lower shock velocities. \\

\textbf{\textit{The West and East regions}}. The spectra of the West and East regions (slit positions 4a, 4b, and 11a) reveal shock-heated gas ([S\,{\sc ii}]/H$\alpha$ $\sim$ 0.76$-$1.51). The lack of [O\,{\sc iii}] emission in these regions suggests a low shock velocity (${\leq}$70 km s$^{-1}$). As seen in Table \ref{Table4}, the East and West regions show no significant differences in spectral properties. \\

Using {\it eROSITA} data, \citet{Ca23} investigated SNRs IC~443 and G189.6$+$3.3 in four spectral regions, namely A, B, C, and D (see Figure 1 in their paper). Even though Region B is the bright external part of the G189.6+3.3 emission, the inner region A appears to be surrounded by a brighter X-ray emission. The C region is located in the bright NE region, and the D region is located in the SW region of IC~443. Their regions A and B correspond to our Regions 7, 8, and 9. The region C corresponds to our Region 3.  The region D corresponds to our Regions 4, and 6. Interestingly, they found that in all the regions, the X-ray spectral analysis revealed a constant temperature ($kT$ = 0.7 keV) together with an almost uniform column density ($N_{\rm H}$ $\sim$ 4$\times10^{21}$ cm$^{-2}$). They concluded that these results support that G189.6+3.3 completely overlaps with IC~443 and G189.6$+$3.3 is in front of IC~443. We estimated column density ($N_{\rm H}$) values only for Region 3, ($\sim$4.8$-$8.9)$\times$ $10^{21}$ cm$^{-2}$, which is one or two times the order of the value of \citet{Ca23}.

\subsection{Atomic environment}
Our H\,{\sc i} analysis newly found a shell-like distribution of atomic gas that seems to be trace the radio-shell boundary of G189.6$+$3.3. Since the radial velocity of the H\,{\sc i} shell is roughly consistent with the systemic velocity of IC~443, these two SNRs are likely located at the same distance. If the scenario is correct, G189.6$+$3.3 and IC~443 may influence each other. According to \cite{2015A&A...580A..49I}, multiple-episodes of atomic gas compression such as stellar winds and/or slow-supernova shocks can naturally produce a converging flow that forms filamentary molecular and/or atomic clouds. The shocked H\,{\sc i} cloud of IC~443 particularly in Fig.~\ref{figure14}(a)--\ref{figure14}(c) shows such a filamentary distribution, which can only be seen within the overlap region of the two SNR shells. The high-velocity wings of H\,{\sc i} toward the overlap region of the two SNRs also provide a strong support of the shock-broadening of H\,{\sc i} spectra due to the converging flow (Fig.~\ref{figurepv}(b)). Further high-resolution H\,{\sc i} studies with a wide-spatial coverage are needed to certify the physical association between the two SNRs and the formation mechanism of H\,{\sc i} filamentary clouds. \\

In summary, our optical observations and  H\,{\sc i} analysis of IC~443 and G189.6$+$3.3 allowed us to better define their properties. The ranges of our estimated electron density and pre-shock cloud density clearly indicate the complex structure surrounding IC~443 and G189.6+3.3. Our H\,{\sc i} analysis suggests that these two SNRs are likely located at the same distance. Future high-resolution X-ray spectra with the {\it XRISM} and {\it Athena} will help us to understand the physical processes of these SNRs in more detail.\\

\section*{Acknowledgements}
We thank the referee for valuable comments and suggestions that helped to improve the paper. We also thank T\"{U}B\.{I}TAK National Observatory (TUG) for partial support in using the T100 telescope with project number 1592 and  RTT150 with project number 1562. The Canadian Galactic Plane Survey (CGPS) is a Canadian project with international partners. The Dominion Radio Astrophysical Observatory is operated as a national facility by the National Research Council of Canada. The CGPS is supported by a grant from the Natural Sciences and Engineering Research Council of Canada. H.S. was also supported by JSPS KAKENHI grant No. 21H01136.

\section*{DATA AVAILABILITY}

The optical data taken with the RTT150 and T100 telescopes will be made available by the corresponding author upon request.

%%% FACILITIES
 
%%%%%%%%%%%%%%%%%%%% REFERENCES %%%%%%%%%%%%%%%%%%

% Don't change these lines
\bsp	% typesetting comment
\label{lastpage}
\end{document}